\documentclass{nature}
\usepackage{lscape}
\usepackage{graphicx}
\usepackage{epstopdf}
\usepackage{graphicx}
\usepackage{multirow}
\usepackage{color}
\usepackage{rotating}
\usepackage{wasysym}
\usepackage{tabularx}
\usepackage{colortbl}
\usepackage{url}
\usepackage{caption}

\newcommand{\eps}{e$^{-}$ s$^{-1}$}
\newcommand{\e}{e$^{-}$}
\newcommand{\nw}{nW m$^{-2}$ sr$^{-1}$}

\title{Measurement of the Cosmic Optical Background using the Long Range
  Reconnaissance Imager on New Horizons}

\author{
Michael Zemcov$^{1,2}$,
Poppy Immel$^{1}$, 
Chi Nguyen$^{1}$, 
Asantha Cooray$^{3}$, 
Carey M.~Lisse$^{4}$, \&
Andrew R.~Poppe$^{5}$
}

\begin{document}

\maketitle

\begin{affiliations}
\item Center for Detectors, School of Physics and Astronomy, Rochester
  Institute of Technology, 1 Lomb Memorial Dr., Rochester NY 14623, USA.
\item Jet Propulsion Laboratory (JPL), 4800 Oak Grove Dr., Pasadena, CA 91109, USA.
\item Department of Physics \& Astronomy, University of California, Irvine, CA 92697, USA.
\item Planetary Exploration Group, Space Department, Johns Hopkins University Applied Physics Laboratory, 11100 Johns Hopkins Rd., Laurel, MD 20723.
\item Space Science Laboratory, University of California at Berkeley, Berkeley,
  CA, 94720, USA.  \\ Correspondence and requests for materials should be
  addressed to M.Z.~(email: zemcov@cfd.rit.edu).
\end{affiliations}

\begin{abstract}
  The cosmic optical background is an important observable that
  constrains energy production in stars and more exotic physical
  processes in the universe, and provides a crucial cosmological
  benchmark against which to judge theories of structure formation.
  Measurement of the absolute brightness of this background is
  complicated by local foregrounds like the Earth's atmosphere and
  sunlight reflected from local interplanetary dust, and large
  discrepancies in the inferred brightness of the optical background
  have resulted.  Observations from probes far from the Earth are not
  affected by these bright foregrounds.  Here we analyze data from the
  Long Range Reconnaissance Imager (LORRI) instrument on NASA's New
  Horizons mission acquired during cruise phase outside the orbit of
  Jupiter, and find a statistical upper limit on the optical
  background's brightness similar to the integrated light from
  galaxies.  We conclude that a carefully performed survey with LORRI
  could yield uncertainties comparable to those from galaxy counting
  measurements.
\end{abstract}



The cosmic optical background (COB) is the summed emission from all
sources outside of our Milky Way galaxy emitted at wavelengths roughly
corresponding to those visible with the human eye.  It is a powerful
diagnostic of the emission from known astrophysical processes in
galaxies including stellar nucleosynthesis, mass accretion onto black
holes, and the gravitational collapse of
stars\cite{Tyson1995,Hauser2001,Cooray2016}.  A comparison of the COB
intensity to the surface brightness arising from known galaxy
populations can reveal the presence of diffuse backgrounds produced by
more exotic phenomena such as the decay of particle species outside
the standard model or light from objects outside of
galaxies\cite{Cooray2012,Zemcov2014,Gong2016}.

Direct photometric measurement of the COB has proven to be
challenging.  The earth's atmosphere is several orders of magnitude
brighter than the COB, and accounting for the various relevant
emission, absorption, and scattering effects is a daunting task.
Sunlight scattered from interplanetary dust (IPD) particles in the
Solar system, known as Zodiacal light (ZL) when viewed from the earth,
also produces a large foreground to direct measurement of the COB from
vantage points in the inner Solar system.  Though progress has been
made in carefully accounting for the atmosphere and ZL in the
optical\cite{Bernstein2007,Mattila2012} and into the
near-IR\cite{Gorjian2000, Wright2001, Cambresy2001, Wright2004,
  Levenson2007,Matsumoto2005}, as it is typically $> 100$ times
brighter than the COB small errors in this accountancy propagate to
large errors on the COB\cite{Mattila2003, Mattila2006}.  It is thus
desirable to measure the COB from vantage points where the earth's
atmosphere and the light from IPD are not appreciable components of
the diffuse sky brightness, such as the outer parts of our Solar
system\cite{Cooray2009}.  Though many planetary probes have had
optical-wavelength cameras, they are rarely designed with the demands
of extragalactic astronomical observations in mind.

Two exceptions to this are the early NASA probes Pioneer 10
and 11, which were instrumented with imaging photopolarimeters (IPPs)
that returned measurements of the sky brightness ranging from 1 to 5.3
AU\cite{Weinberg1974}.  These data have been used to measure both the
decrease in the IPD light with heliocentric distance\cite{Hanner1974},
diffuse light from the Galaxy\cite{Toller1987, Gordon1998}, and the
brightness of the COB itself\cite{Toller1983, Matsuoka2011} using the
two IPP bands spanning $390{-}500 \,$nm and $600{-}720 \,$nm.  The
Pioneer measurements remain the most stringent constraints of
the COB\cite{Matsuoka2011}, and have uncertainties dominated by errors
associated with subtracting galactic components including the
integrated light from stars (ISL) and diffuse galactic light (DGL).

NASA's New Horizons spacecraft\cite{Weaver2008} recently performed the
first detailed reconnaissance of the Pluto-Charon system.  It includes
as part of its instrument package the Long Range Reconnaissance
Imager\cite{Conard2005, Morgan2005, Cheng2008} (LORRI), an optical
camera with sensitivity over a broad $440{-}870 \,$nm half-sensitivity
passband.  Importantly, rather than a scanning photometer like the
IPP, LORRI is a Newtonian telescope with characteristics including
excellent pointing stability, a $20.8 \,$cm diameter
Ritchey-Chr{\'e}tien telescope, an $0^{\circ}.3 \times 0^{\circ}.3$
instantaneous field of view,
$1^{\prime \prime} \times 1^{\prime \prime}$ pixels, and (crucially)
real-time dark current monitoring.  The achieved point source
sensitivity of LORRI is $V=17$ in a $10 \,$s exposure in $4 \times 4$
pixel on-chip ``rebinning'' mode, making it a sensitive astronomical
instrument.  As a result of this sensitivity and angular resolution,
much of the starlight that challenged the earlier Pioneer
measurements can be resolved out in LORRI images, providing a
relatively clean measurement of diffuse astrophysical emission.  

In this paper, we use archival data from the New Horizons checkout and
cruise phases to measure the COB from several vantage points in the
Solar system.  We correct for dark current in the detectors, mask
bright stars from the images, assess the amplitude of residual
starlight, sunlight from interplanetary dust, and diffuse galactic
light, and correct for galactic extinction to measure
$\lambda I_{\lambda}^{\rm COB} = 4.7 \pm 7.3 (\mathrm{stat.})$
$^{+10.3}_{-11.6} (\mathrm{sys.}) \,$\nw, giving a $2 \sigma$
statistical upper limit of
$\lambda I_{\lambda}^{\rm COB} < 19.3 \,$\nw, which excludes some of
the early results in the literature.  This measurement is based on a
very limited data set with characteristics that complicate
astrophysical examination.  We conclude that a carefully designed
survey of the COB from LORRI beyond the orbit of Pluto has the
potential of definitively measuring its surface brightness away from
the complicating effects of the earth's local interplanetary dust
cloud.

\section*{Results}
\label{S:results}
\vspace*{-15pt}

\paragraph{Data Set.}  In this study we concentrate on the
$4 \times 4$ pixel ``rebinned'' LORRI exposures, for which on-chip
summing has been used to improve the surface brightness sensitivity
over the native-resolution data.  This rebinning mode is particularly
advantageous in the small signal regime where the read noise penalty
is large.  The rebinned images have spatial resolution of
$4^{\prime \prime}.3$ over a full $256 \times 256$ (super-)pixel
frame.  We term magnitudes in this LORRI band $R_{\rm L}$, as it is
close to (though much broader than) the Johnson-Cousins $R$-band at
$640 \,$nm\cite{Bessell1998}; in fact the LORRI bandpass covers
essentially all of astronomical $V$, $R$, and $I$ bands.

\begin{figure*}[ht!]
\centering
\includegraphics*[height=2.96in]{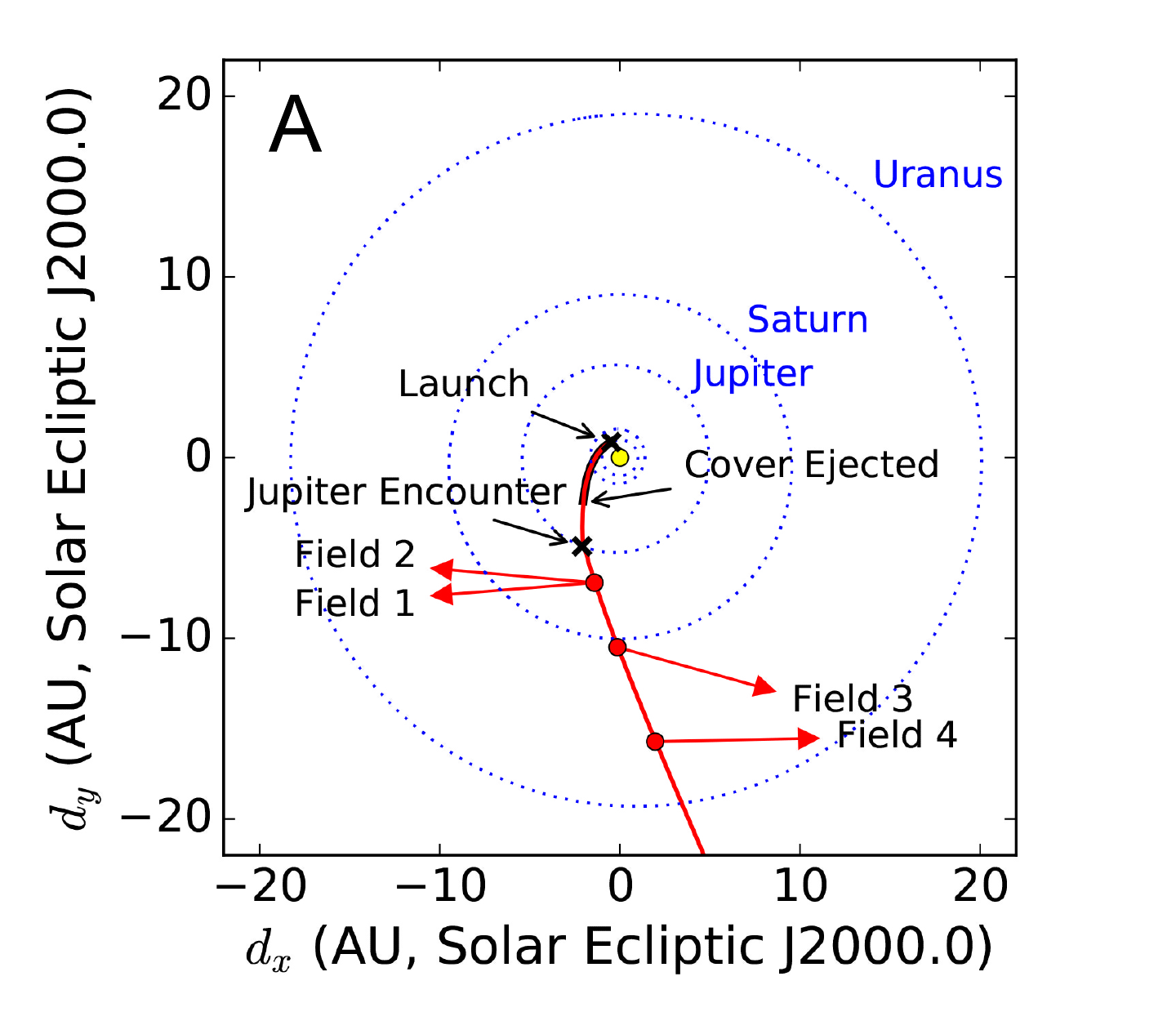}
\includegraphics*[height=2.96in]{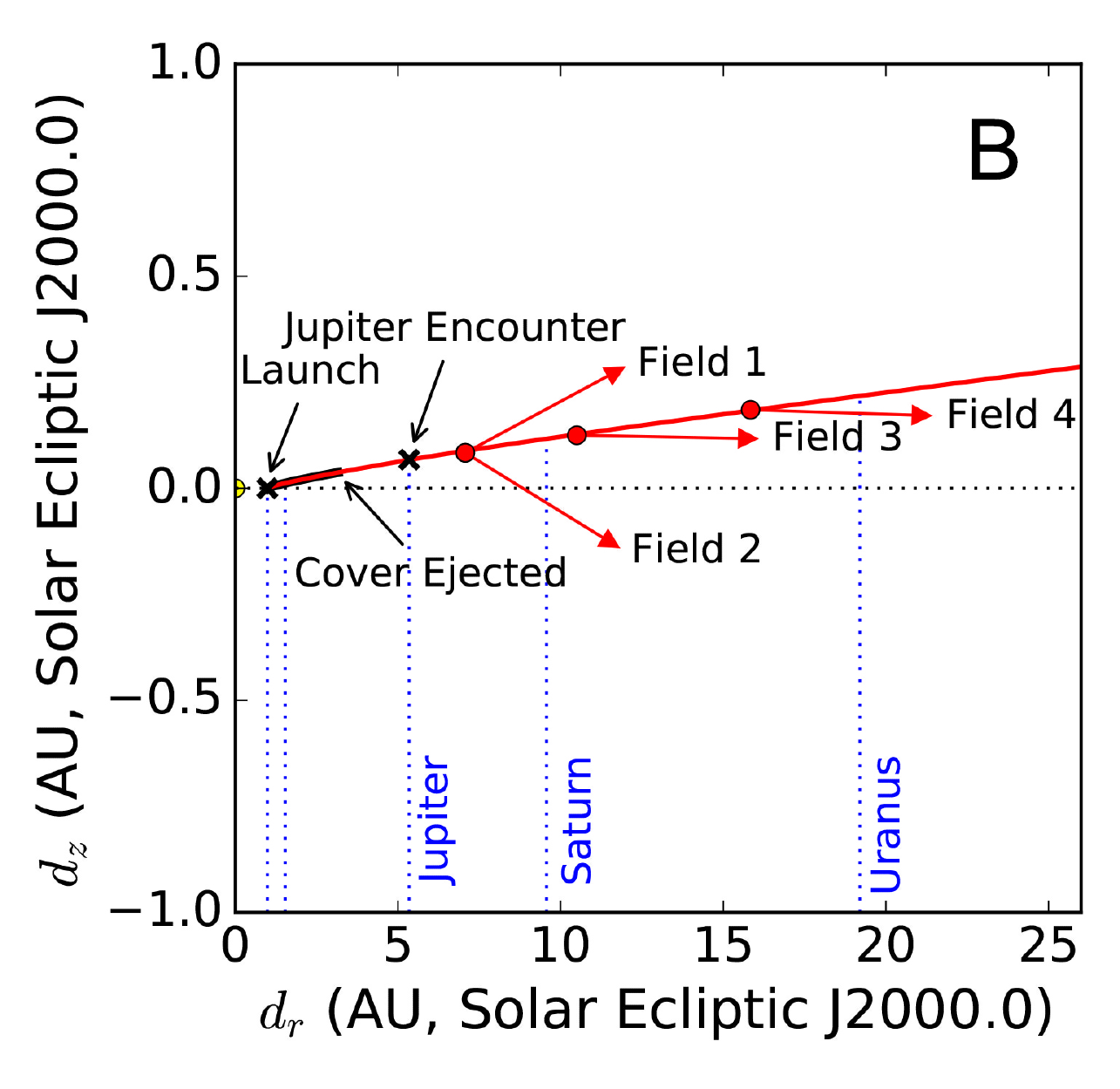}
\caption{\textbf{The trajectory of New Horizons through the solar
    system.}  Data collection periods of relevance to this study are
  indicated.  Both the $x-y$ and $r-z$ planes are shown (panels A and
  B, respectively), with the axes in solar ecliptic units and
  $d_{r} = \sqrt{d_{x}^{2} + d_{y}^{2}}$.  New Horizons was launched
  from Earth at 1 AU, and data with the LORRI dust cover in place were
  acquired at 1.9 AU, just beyond Mars' orbit at 1.5 AU (inner blue
  dotted lines).  The dust cover was ejected near 3.6 AU, and data
  were acquired prior to and during an encounter with Jupiter.  The
  data considered here were taken between 2007 and 2010 while New
  Horizons was in cruise phase.  The red vectors indicate the relative
  positions of fields $1{-}4$ compared to the sun and plane of the
  ecliptic.  \label{fig:trajectory} }
\end{figure*}

New Horizons was launched January 19, 2006.  The path of
New Horizons through the Solar system is summarized in Figure
\ref{fig:trajectory}.  Approximately 90 days after launch, some 359
dark data sets were acquired by LORRI while approximately 1.9 AU from
the sun.  During this time, the LORRI dust cover was in place over the
telescope aperture, providing close to optically dark conditions.  The
cover was ejected on 2006 August 29, and shortly thereafter a series
of images of the open star cluster Messier 7 were acquired.  From
these data, the New Horizons team determined a preliminary
photometric calibration, a full width at half maximum (FWHM) pointing
jitter of $0.45$ pixels, and geometric distortion less than $0.2$
pixels across the field of view\cite{Cheng2008}.  Following this, a
series of short exposure test images was acquired, and at the
beginning of 2007 the first science image was taken of Callirrhoe, a
small irregular moon of Jupiter, with a full $10 \,$s exposure time.
A series of images of Jupiter and its immediate environment were then
acquired during an encounter, which we do not consider here.  Closest
approach to Jupiter occurred on 2007 February 28.  Following the
Jupiter encounter, New Horizons entered cruise phase.  LORRI
data were acquired on an approximately annual basis, and consisted of
$10 \,$s observations of distant Solar system objects.  The current
public archival data records end 2014 July 20, approximately a year
prior to the Pluto encounter.

We cut data using requirements on integration time, solar elongation,
and thermal dust emission, following which we are left with fields
$1{-}4$ whose characteristics are summarized in Table \ref{tab:gooddata}.  

\begin{center}
\begin{minipage}{\linewidth}
\begin{center}
\captionof{table}{Data sets used in this analysis. \label{tab:gooddata}}
\begin{tabular}{c|cc|cc|c|c}
\hline
Field Number & $\alpha$ (J2000)& $\delta$ (J2000) & $\ell$ & $b$ & $e_{b}$ & $A_{V}$ \\
& hh:mm:ss & hh:mm:ss &  ($^{\circ}$) & ($^{\circ}$) & ($^{\circ}$) & mag \\ \hline

1 & 13:04:02 & 23:57:02 & 345.4147 & 85.7384 & 28.2096 & 0.06 \\

2 & 10:47:36 & $-26$:46:56 & 271.4532 & 28.4141 & $-31.5843$ & 0.22 \\

3 & 23:04:27 & $-7$:07:00 & 66.2722 & $-57.6861$ & $-1.0847$ & 0.16 \\

4 & 00:07:14 & $-1$:15:00 & 98.8079 & $-62.0328$ & $-1.8651$ & 0.10 \\

\hline
\end{tabular}
\end{center}
\end{minipage}
\end{center}

\paragraph{COB measurement.}  The brightness in an arbitrary
image of the astronomical sky acquired above the Earth's atmosphere
$\lambda I_{\lambda}^{\rm meas}$ can be expressed as:
\begin{equation}
\lambda I_{\lambda}^{\rm meas} = \lambda I_{\lambda}^{\rm IPD} +
\lambda I_{\lambda}^{\ast} + \lambda I_{\lambda}^{\rm RS} + \lambda
I_{\lambda}^{\rm DGL} + \epsilon \lambda I_{\lambda}^{\rm COB} + \lambda
I_{\lambda}^{\rm inst},
\end{equation}
where $\lambda I_{\lambda}^{\rm IPD}$ is the brightness associated
with interplanetary dust, $\lambda I_{\lambda}^{\ast}$ is the
brightness associated with resolved stars,
$\lambda I_{\lambda}^{\rm RS}$ is the brightness associated with
residual starlight from stars too faint to be detected individually or
the faint wings of masked sources, $\lambda I_{\lambda}^{\rm DGL}$ is
the brightness of the DGL, $\lambda I_{\lambda}^{\rm COB}$ is the
brightness of the COB, $\epsilon$ is a factor accounting for
absorption in galactic dust, and $I_{\lambda}^{\rm inst}$ is brightness
associated with the instrument including all potential contributions
to the measured zero-point offset.  The major difficulty with COB
measurements is that, with the exception of
$\lambda I_{\lambda}^{\ast}$, each of these sources can have an
isotropic component, which is problematic since the COB itself is
isotropic.  As a result, care must be taken to understand and correct
for the brightness of each component, particularly those that appear
constant over angular scales similar to the field of view of the
instrument.

\begin{figure*}[p]
\centering
\includegraphics*[width=6.5in]{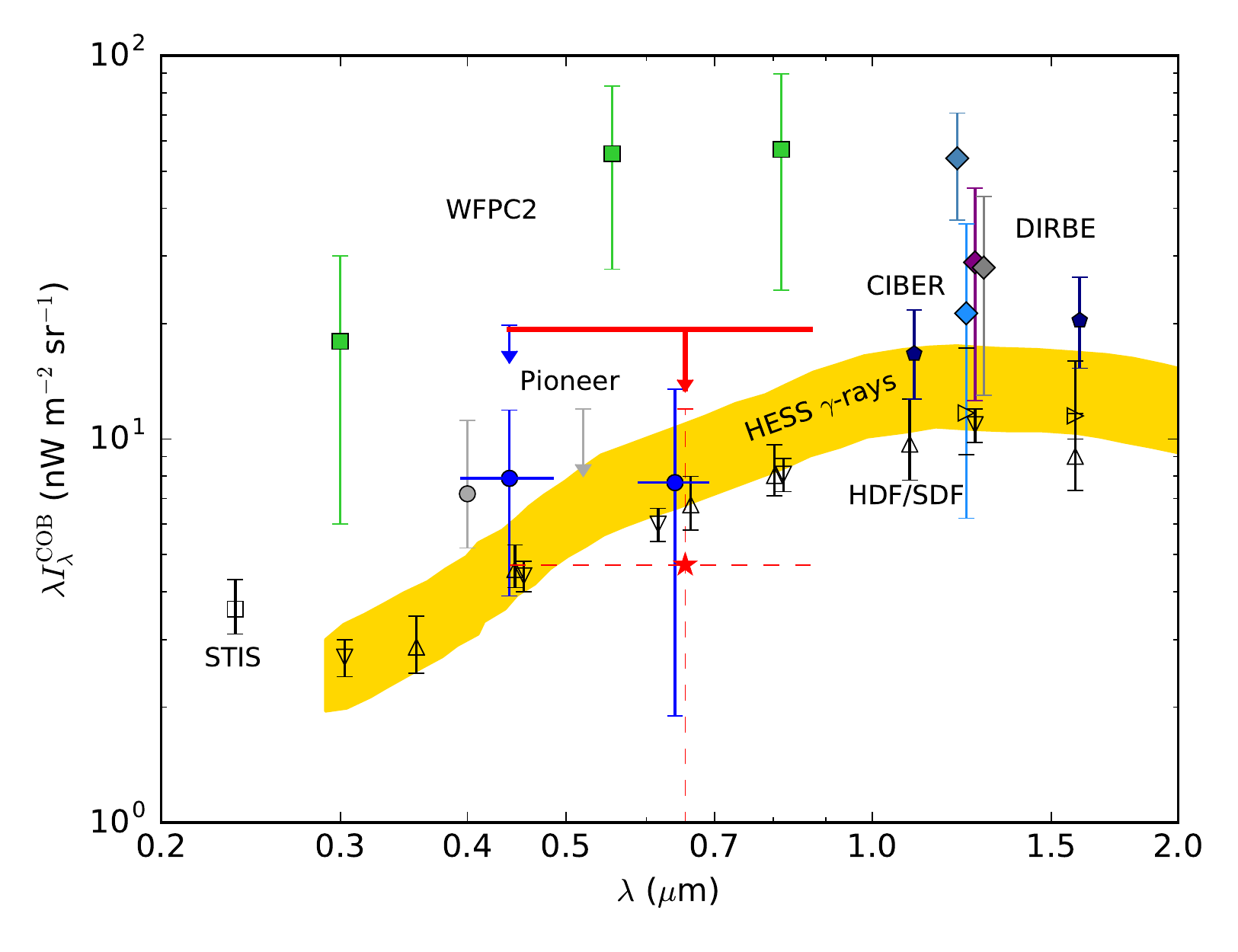}
\caption{\textbf{Measurements of the COB surface brightness.} The
  $\lambda I_{\lambda}^{\rm COB}$ determined in this study are shown
  as both an upper limit (red) and a mean (red star).  We also show
  previous results in the literature, including direct contraints on
  the COB (filled symbols) and the IGL (open symbols).  The plotted
  LORRI errors are purely statistical and are calculated from the
  observed variance in the mean of individual $10 \,$s exposures; see
  Figure \ref{fig:systematics} for an assessment of the systematic
  uncertainties in the measurement.  We include the measurements from
  HST-WFPC2\cite{Bernstein2007} (green squares), combinations of DIRBE
  and 2MASS\cite{Wright2001,Cambresy2001,Wright2004,Levenson2007}
  (diamonds; the wavelengths of these measurements have been shifted
  for clarity), a measurement using the ``dark cloud''
  method\cite{Mattila2012} (grey circles), and previous Pioneer 10/11
  measurements\cite{Toller1983,Matsuoka2011} (blue upper limit leader
  and circles).  The gold region indicates the H.E.S.S.~constraints on
  the extragalactic background light\cite{HESS2013}.  We include the
  background inferred from CIBER\cite{Zemcov2014} (pentagons).  The
  IGL points are compiled from HST-STIS in the UV \cite{Gardner2000}
  (open square), and the Hubble Deep Field\cite{Madau2000} (downward
  open triangles) the Subaru Deep Field\cite{Totani2001, Keenan2010}
  (upward open triangles and sideways pointing triangles) in the
  optical/near-IR.  Where plotted, horizontal bars indicate the
  effective wavelength band of the measurement.  Our new LORRI value
  from just $260 \,$s of integration time is consistent with the
  previous Pioneer values. \label{fig:cob} }
\end{figure*}

We isolate $\lambda I_{\lambda}^{\rm COB}$ using three basic steps:
mask stars near or brighter than the detection threshold to remove the
effect of $\lambda I_{\lambda}^{\ast}$; subtract the diffuse
components either originating in the instrument or from local
astrophysical emission to isolate the diffuse residual component
$\lambda I_{\lambda}^{\rm resid} = \epsilon \lambda I_{\lambda}^{\rm
  COB}$;
and correct the mean residual intensity for the effects of galactic
extinction to yield $\lambda I_{\lambda}^{\rm COB}$.  Averaging over
all the fields using inverse noise variance weighting, we determine
that $\lambda I_{\lambda}^{\rm COB}= 4.7 \pm 7.3 \,$\nw, where the
uncertainty is purely statistical and is assessed from the scatter in
the individual exposures.  This gives a $2 \sigma$ upper limit on the
COB brightness of
$\lambda I_{\lambda}^{\rm COB}(2 \sigma) < 19.3 \,$\nw.  Our
measurement and comparisons with previous measurements in the
literature are shown in Figure \ref{fig:cob}.

\begin{figure*}[p]
\centering
\includegraphics*[width=6.5in]{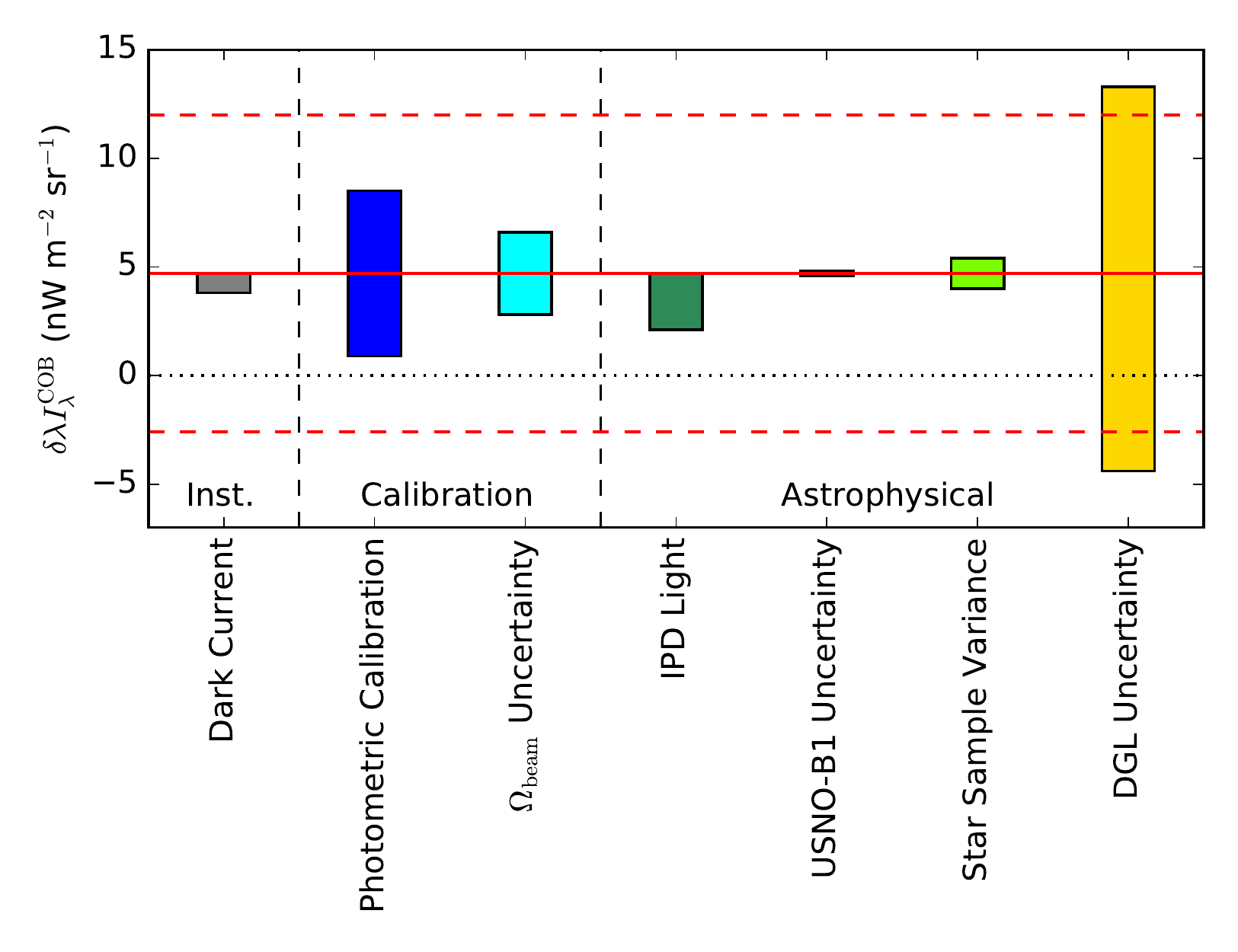}
\caption{\textbf{Summary of the various systematic errors in our
    determination of $\mathbf{\lambda I_{\lambda}^{\rm \bf COB}}$.}  The various
  sources of uncertainty are labeled, with the colored bars showing
  their variation from the mean value we measure (red solid lines; see
  also Supplementary Table 3).  Most of the errors are smaller than
  the statistical uncertainty of the measurement (dashed red lines),
  except for the uncertainty in the DGL model which is large compared
  to the other errors.  We do not show the errors associated with the
  optical ghosts and extinction correction as these are substantially
  less than a significant figure.  The dominant uncertainties in this
  measurement are in fact not statistical, and to a great extent
  depend on the fields chosen and ancillary data available, so further
  observations in a dedicated survey program hold great
  promise.  \label{fig:systematics} }
\end{figure*}

This measurement is also subject to various systematic uncertainties
associated with the calibration and foreground removal.  We carefully
assess these errors by probing the allowed variation in each of the
models and measurements to derive an overall calibration and
systematic uncertainty budget, summarized in Figure
\ref{fig:systematics}.  As the astrophysical errors identified in this
analysis are uncorrelated, combining them in quadrature is an
appropriate estimate of the total error present in the measurement.
This is not the case for the calibration errors, which we add linearly
and then sum in quadrature with the astrophysical foregrounds to give
a conservative total systematic error estimate of $\{+10.3,-11.6\} \,$\nw.

\section*{Discussion}
\label{S:discussion}
\vspace*{-15pt} These data show the power of LORRI for precise,
low-foreground measurements of the COB.  The measurement presented
here is not consistent with the earlier HST-WFPC2
constraints\cite{Bernstein2007}, but is consistent with both the
Pioneer\cite{Matsuoka2011} and ``dark cloud\cite{Mattila2012}''
measurements, as well as the $\gamma$-ray inference\cite{HESS2013}.
This measurement constrains the possibility of a COB significantly in
excess of the expectation from IGL.

Though the bandwidth required to telemeter the data from the outer
Solar system constrains the number of observations possible, with a
carefully designed survey we should be able to produce a definitive
measurement of the diffuse light in the local universe, and a tight
constraint on the light from galaxies in the optical wavebands.
LORRI's ability to resolve much of the starlight has significantly
reduced the potential for foreground contamination compared to
measurements from the Pioneer IPP, and the LORRI field is
small enough that bespoke ground-based assessments of the faint
starlight in each field are conceivable.  As a result, a future LORRI
survey would benefit from careful design and \textit{pre facto}
observations of the survey fields.  Given the total integration time
used in this measurement was only $260 \,$s, a total integration time
of $\sim 4.5 \,$hrs would allow us to achieve $\sim 1 \,$\nw\
statistical uncertainties.  Because LORRI can allow $30 \,$s
integrations, this hypothetical measurement would require $\sim 500$
integrations, which is not prohibitive in terms of data storage
nor telemetry requirements.  

It would be particularly useful to observe high galactic latitude
fields at a variety of ecliptic latitudes and solar elongations to
search for IPD light.  Though likely to be too faint to detect, models
suggest there may be an increase in the IPD population towards the
Kuiper Belt from collisional material\cite{Poppe2016}.  This increase
may be observable in the IPD light intensity with a carefully
designed, deep survey.  At the very least, observations of the inner
Solar system from New Horizons' perspective may provide
useful new information about the global structure of the IPD cloud.
In addition, currently unpublished instrument calibration information
such as susceptibility to off-axis light and detailed pointing
stability assessments could improve the accuracy of this kind of
measurement.

A primary lesson learned from this analysis is that, following the
accurate removal of ISL, the DGL estimate becomes the largest source of
uncertainty.  
Because of the uncertainty in the measured $100 \, \mu$m background
level, the DGL-$I_{100 \, \mu \mathrm{m}}$ scaling, and the galactic
latitude dependence of the scattering, this component varies by
approximately the expected brightness of the IGL in each of our
fields.  For example, Field 1 is very close to the galactic pole where
the DGL should be faint, but even here the models and observations
suggest the DGL brightness could vary from $4$ to $11 \,$\nw\ at one
standard deviation.  Future observations with LORRI should concentrate
on the lowest $I_{100 \, \mu {\rm m}}$ fields available on the sky to
minimize the uncertainty.  If many statistically independent fields
are sampled, the DGL-$I_{100 \, \mu {\rm m}}$ linear regression
technique we briefly explore should permit measurement of the
optical-thermal infrared correlation precise enough to allow sub-\nw\
determination of the COB.  Improved DGL characterization using other
techniques are also of continuing importance.

This measurement of the COB brightness, while not currently as precise
as those from Pioneer\cite{Matsuoka2011}, is important as it
suffers from completely different instrumental and foreground
uncertainties as the existing measurement.  It is also the only
measurement sensitive to the $I$-band $700{-}900$nm wavelength range.
Though some challenges remain, further data from LORRI could provide a
definitive measurement of the extragalactic background light at
optical wavelengths, and may be instrumental in completing our
understanding of the history of stars and galaxies in the universe.

\section*{Methods}
\vspace*{-15pt}
\paragraph{Observational Data.}

The basic New Horizons flight timeline is given in
Supplementary Table 1, including a summary of the
data taken during the checkout and cruise phases.  The LORRI bandpass
has an effective wavelength of $\lambda = 655 \,$nm for a
flat-spectrum source, with half power response from $440{-}870 \,$nm.
Supplementary Table 1 indicates the dates of the
observations, the notional targets, the number of integrations
available, the exposure time per integration, as well as astrophysical
information like the heliocentric distance $R_{\odot}$, solar
elongation $\theta_{\odot}$, and the $100 \, \mu$m specific intensity
$I_{100 \mu {\rm m}}$.  In this work, we refer to each data set with a
field number, running over D$1{-}4$, R$1{-}10$, and $1{-}4$, using the
numbering scheme presented below.

We performed cuts on the full data set to account for various factors
affecting the data quality.  First, we restricted our attention to the
data with integration times $t_{\rm int} > 1 \,$s, eliminating sets
R$1{-}4$.  Second, because the large-angle response of the LORRI
telescope shows response from diffuse scattering of sunlight
illuminating the light baffle\cite{Cheng2010}, we
remove data for which $\theta_{\odot} < 90^{\circ}$.  Finally, many of
these LORRI data sets are taken at low galactic latitudes where the
DGL is bright.  We exclude data sets for which
$I_{100 \mu {\rm m}} > 10 \,$MJy sr$^{-1}$, which removes fields
R${5}$ and R$7{-}10$ from analysis.  Though excluding much of the most
useful data is not ideal, these fields are measured very close to the
Galactic plane where the contamination from the local environment
precludes careful measurement of the faint signals from either the COB
or interplanetary dust in reflection.

\paragraph{Data Reduction.}
The archival LORRI data are available in a format in which they have
already been processed through several instrument calibration steps
including bias correction, smear correction, and relative pixel
response correction\cite{Cheng2008,NHICD}.  In brief, the raw data
consists of voltages measured at the end of the exposure reported in
data numbers (DN). In the first step of the processing, the median of
the dark reference pixels is used to subtract the global reference
voltage, and a reference ``super-bias'' frame measured during ground
testing is used to correct for bias variations over the array.  Next,
image smearing and flat field corrections measured during ground
calibration are used to account for image smearing and relative pixel
response.  The files that are input to our processing have units data
number per integration time and, though they have been partially
processed, must be: astrometrically registered; masked for known
detector defects, transients, and bright astronomical objects;
corrected for instrumental effects; and calibrated to photometric
units.  Example images of the four science fields in this study are
shown in Supplementary Fig.~1, calibrated to
$\lambda I_{\lambda}$.

\paragraph{Astrometric Registration.}
Astrometric registration is required to allow masking of bright stars,
which is necessary to account for $\lambda I_{\lambda}^{\ast}$.
Functionally, this means we want to ensure that each pixel in an image
is accurately associated with a pair of right ascension and
declination coordinates, $(\alpha, \delta)$.  We determine the
orientation and the scale of each image using the publicly-available
astrometric calibration software package
\url{http://astrometry.net}\cite{Lang2010}.  The algorithm uses a
four-step procedure in which: bright sources in each
image are detected; the detected sources are divided
into subsets whose relative positions are recorded and matched against
a pre-built index; the solution is verified using
predictive star position checks; and, the final
alignment information is returned to the user in a Flexible Image
Transport System ({\sc fits}) header for each file.  For the LORRI
data, this algorithm successfully solved the astrometric registration
for each field independently.  Further, constant parameters of the
instrument like the pixel size and image distortions are found to be
consistent between observations over the entire data set, with very
small uncertainties.  The registration information returned by the
software package is used to calculate $(\alpha, \delta)$ for every
pixel in each image.

\paragraph{Masking.}
In order to reliably measure the diffuse sky brightness, it is
necessary to exclude residual instrumental signal and detectable point
sources from the images that contribute to
$\lambda I_{\lambda}^{\rm inst} $ and $\lambda I_{\lambda}^{\ast} $,
respectively.  We implement image masks to remove brightness
associated with: stars near or brighter than the detection limit;
pixels that may suffer from electronic or optical pathologies; and
cosmic rays and other transient events.

To mask stars, we use the USNO-B1 catalog\cite{Monet2003}, which
provides photometric fluxes in approximately Johnson-Cousins $B$, $R$
and $I$ bands over the entire sky.  Though in some regions this
catalog reaches completeness of $V=21$, it is nonuniform and source
fluxes are calibrated to only 0.25 mag accuracy.  We synthesize the
$R_{\rm L}$ flux by fitting a linear model to the USNO-B1 measurements
and compute the LORRI band-weighted integral of the flux for each
source.  To compromise between maximal removal of stars and minimal
removal of galaxies which contribute to the COB, we mask the fields at
a flux limit of $R_{\rm L}=17.75$, which is $\sim 1 \,$mag below the
$1 \sigma$ point source sensitivity in the images.  This threshold has
the added benefit of moving the mask flux threshold away from the
USNO-B1 catalog completeness limit where the survey uniformity is
problematic.  Given LORRI's small field of view, we calculate that the
error introduced in the final COB estimates due to accidentally
masking galaxies at the bright end of the number
counts\cite{Driver2016} is 0.006\nw.  A search of the available
optical data in these fields is consistent with the number counts, and
we find that no exceptionally bright galaxies fall into these fields.

To build an appropriate source mask, we also require accurate
knowledge of the instrument point spread function (PSF).  We use a
stacking method\cite{Bock2013} to sum the emission from all
$R_{\rm L}<16$ sources in the image to form an estimate of the LORRI
PSF.  Briefly, for each source brighter than the magnitude limit, we
interpolate the image onto a ten times finer grid centered on the
cataloged source position, and sum all such postage-stamp images. In
each postage-stamp image, pixels far from any star images or masked
pixels are used to calculate the zero point of the image.  No lower
magnitude limit is required as none of the images contain stars bright
enough to induce non-linear response in the detector.  The stacked PSF
is shown in Supplementary Fig.~2.  We compute the uncertainty in the
PSF by performing two stacks, one stacking on a random half of the
sources in the catalog and the other on the other half.  We find a
FHWM of $1.53\pm 0.05^{\prime \prime}$, consistent with both
laboratory\cite{Morgan2005} and in-flight\cite{Owen2012}
measurements of the PSF of $1.5^{\prime \prime}$.

We mask stars in each LORRI image by using the band-weighted magnitude
estimate $m$ to calculate a radius around each source to exclude from
analysis.  This radius $r(m)$ is computed from:
\begin{equation}
r(m) = 2.5 \left( \frac{m_{\rm lim}}{m} \right)^{2},
\end{equation}
where the free parameters are determined empirically from the data.
Here, $m_{\rm lim} = 17.7$ at $R_{\rm L}$-band and $r(m)$ has units of
pixels.  To assess the efficacy of this mask as a function of
magnitude, we simulate noiseless images of the stars in each field per
magnitude bin, apply the mask, and then calculate the residual surface
brightness.  We find the largest contribution to the residual
brightness is from stars near the limiting magnitude, which contribute
at most $0.15 \,$pW m$^{-2}$ sr$^{-1}$ per source.  Brighter sources
have larger masks and contribute less total surface brightness as there
are fewer of them.  We calculate the total flux left in the images
from residual unmasked star flux as part of the ISL assessment.

The data used in our COB study have a variety of Solar system objects
as their primary targets.  Though Haumea and Makemake should be faint
in the optical ($R > 15$), even at a distance of $23.2 \,$AU Neptune
appears bright in the LORRI images ($R \sim 7$).  To account for this,
we uniformly mask the central $2^{\prime}.3 \times 2^{\prime}.3$ from
each of the science images.  In the case of Neptune, this corresponds
to $35 \,$R$_{\neptune}$, which is significantly beyond the outermost
known ring at $2.6 \,$R$_{\neptune}$ and the brightest moons
(including Triton at $14.3 \,$R$_{\neptune}$).  The Neptunian system
does extend further, with a small moon orbiting
$\sim 2000 \,$R$_{\neptune}$ (at this distance $2^{\circ}.2$) from the
planet, and we cannot exclude the possibility that a dust halo far
from the planet is reflecting sunlight and increasing the surface
brightness in the LORRI image.  We do not observe residual structure
in these images, so any such contamination from a circum-Neptunian
dust cloud is relatively faint.  In principle, Haumea and Makemake may
have their own dust clouds, and we cannot exclude the possibility from
these data.  As a result of these considerations, formally our
measurements must be considered to be upper limits to the surface
brightness of the COB.  Future observations away from known Solar
system objects would be beneficial in this regard.

For the Neptune field, the image of the planet is bright enough to
induce charge transfer artifacts in the detector, so we
mask three pixel wide stripes in both the vertical and horizontal
directions of each image.

LORRI is known to have optical ghosting from reflections in the field
flattening lens group for sources that fall between just off the field
to up to $0^{\circ}.37$ from the field\cite{Cheng2008}.  This ghost
is visible in the field 1 images, but not in the other fields.  The
central source mask removes a large fraction of this emission, but we
also manually mask the ghosts in the field 1 images.

To reduce contamination from known defects, we mask both the outermost
five pixels in the image, as well as pixels that are consistently
non-responsive or saturated in the images.  Finally, we apply a clip
mask which excludes pixels $>3 \sigma$ from the mean value of each
image.  This excludes pixels with transient contaminants like
electrical or digitization error and cosmic rays.  On average, we
exclude only $\sim 100$ pixels in this $\sigma$-clipping step,
corresponding to a $0.15 \,$\% loss.  In Supplementary Fig.~3 we
show the same example images shown in Supplementary Fig.~1, but with
the full image mask imposed.

\paragraph{Dark Current and Reference Pixel Behavior.}
Since it cannot be separated from astronomically sourced photocurrent,
an important potential contaminant in this measurement is the dark
current of the detector, which contributes as an approximately
isotropic component of $\lambda I_{\lambda}^{\rm inst}$.  The
operating temperature of the LORRI charge coupled device (CCD) is
$\sim 200 \,$K, so based on the performance of similar devices we
might expect the dark current to be negligible.  However, the COB
measurement is more robust if the dark current contribution can be
completely characterized.

An important feature of the LORRI CCD
is the presence of rows of $4 \times 1024$ (or $1 \times 256$ in rebinned
mode) dark pixels.  These are optically active pixels that are
shielded from incident light by means of a metal lip, but are
otherwise identical to the optically active pixels.  These pixels are
used in the LORRI data reduction pipeline to remove the combination of
dark current and voltage bias in the images.  In our study, these have
the added benefit of giving a fixed reference of the detector array
performance.  

To characterize the long-term performance and stability of the
detector, we compare the measurements performed with the dust cover in
place against those taken with the dust cover off.  The photocurrent
in the reference pixels should be solely a function of the temperature
of the detector\cite{Janesick2001}, which is shown as a function of
time in Supplementary Fig.~4.  During post-launch operations before
the cover was ejected, the detector system was passively cooled to its
final operating temperature of $\sim 193 \,$K.  As a result, the
cover-on data were acquired at a significantly higher temperature than
the optical data.

The detector manufacturer has empirically determined that, in
equal-integration time exposures, the dark
current at a temperature $T$ can be estimated from:
\begin{equation}
\label{eq:idark}
i(T) = 122 \, i_{0} \, T^{3} \exp(-6400/T)
\end{equation}
with $i_{0} \sim 10^{4} \,$\eps pixel$^{-1}$.  From this, we would
predict $i(220 \, {\rm K}) \sim 2.8 \,$\eps pixel$^{-1}$ and
$i(193 \, {\rm K}) \sim 0.035 \,$\eps pixel$^{-1}$.  In Supplementary
Figure 4 we show the mean of the 256 reference pixels in each of the
four cover-on and four cover-off data sets.  These show a decrease
with temperature from a value of $\sim 545 \,$DN to $\sim 542 \,$DN.
Assuming a model in which the bias voltage is a steady-state value
where the dark current is negligible, we also plot both Equation
\ref{eq:idark} and a free-amplitude fit of the same equation in the
figure.  Given the instrumental gain of 22 \e/DN, the free-amplitude
fit gives a mean dark current of $7 \,$\eps\ pixel$^{-1}$ at
$220 \,$K, which is a factor of 2.5 larger than the expectation but
within the manufacturer's expected device to device variance.
Assuming this factor, the expected dark current at $193 \,$K is
$0.09 \,$\eps\ pixel$^{-1}$.

The measured reference bias offset is subtracted from the images as
part of the data processing pipeline.  The value subtracted is the
median of the 256 reference pixels\cite{NHICD}, which is a
reasonable estimate.  However, on closer investigation we find that
the median of the reference pixels can be biased due to the presence
of large outlying pixels from \textit{e.g.} cosmic ray hits.  As a
result, we measure a statistically significant correlation between the
reference row median and the mean of the processed images.  To correct
for this bias, we instead use the $\sigma$-clipped mean of the
reference row for reference subtraction.  The mean is estimated by
rejecting reference pixels with values $> 3 \sigma$ after two
iterations of rejection.  Because in these data the median reference
row value has already been removed, we correct the mean value of the
images by first adding back the reference row median and then
subtracting the $\sigma$-clipped mean value.  The correction is small,
typically $< 0.1 \,$ DN s$^{-1}$.  This procedure effectively
removes the correlation between the subtracted reference value and the
mean value of the processed image.

\paragraph{Photometric Calibration.}
We calibrate the images from DN s$^{-1}$ to Jy pixel$^{-1}$ using
aperture photometry.  For each field, we identify two stars with flux
$< 1000 \,$DN s$^{-1}$ to avoid saturation effects, and greater than
four pixels away from other sources or array artifacts. The pixel
values are summed across a six pixel-wide aperture, and the background
in a three pixel-wide ring three pixels away from the inner aperture
excluding masked pixels is calculated and subtracted.  The
background-corrected aperture sum is then divided by the exposure
time, giving the source flux $S$ in DN s$^{-1}$.  Synthetic photometry
is used to determine the magnitudes in the LORRI band using the
USNO-B1 catalog. We calculate that the reference magnitude at
$R_{{\rm L},0}$ given by the equation:
\begin{equation}
R_{\rm L} = -2.5 \log S + R_{{\rm L}, 0}
\end{equation}
is $18.52 \pm 0.08$. The calibration factors are summarized in
Supplementary Fig.~5.

The conversion from flux to surface brightness relies on both the
frequency of the measurement $\nu_{0}$ and the measured solid angle of
the PSF $\Omega_{\rm b}$ through
$\lambda I_{\lambda} = \nu F_{\nu} / \Omega_{\rm b}$, where
$\Omega_{\rm b}$ is the instrument's 2D image-space impulse response
function integrated over both dimensions.  We estimate the effective
frequency using the measured LORRI passband and assuming a flat input
spectrum, which yields $\nu_{0} = 458 \,$THz.  $\Omega_{\rm b}$ is
calculated by summing the PSF shown in Supplementary Figure
2 over the full $1^{\prime}.8 \times 1^{\prime}.8$ image.
We find $\Omega_{\rm b}=2.64^{+0.18}_{-0.16} \,$pix$^{2}$, in good
agreement with the FWHM$=1.5 \,$pix Gaussian model prediction of
$\Omega_{\rm b} = 2.54 \,$pix$^{2}$.

We estimate the final surface brightness calibration factor to be
$118 \pm 9 \, \mu$Jy/(DN s$^{-1}$), which corresponds to
$50.9 \pm 3.7 \,$\nw/(DN s$^{-1}$).  Following multiplication by this
factor, the images are calibrated in surface brightness units and have
associated masks that can be used to exclude pixels containing
$R < 17.7$ point sources.  The unmasked pixels in these images can be
used to estimate the diffuse sky brightness.  The raw diffuse sky
brightness measurements corresponding to
$\lambda I_{\lambda}^{\rm diffuse} = \lambda I_{\lambda}^{\rm IPD} +
\lambda I_{\lambda}^{\rm RS} + \lambda I_{\lambda}^{\rm DGL} +
\epsilon \lambda I_{\lambda}^{\rm COB}$
are shown in Supplementary Fig.~6.  In order to isolate
the residual component of the observed emission associated with the
COB, it is necessary to account for more local sources of emission,
including residual interplanetary dust, residual starlight, and
diffuse galactic light.  The contribution from each component is
summarized in Supplementary Table 2.

\paragraph{Interplanetary Dust.}
The population of $\sim 1 {-} 1000 \, \mu$m dust particles in the
Solar system reflects light from the sun and sources a diffuse sky
brightness.  Early \textit{in situ} measurements of the dust
distribution in the inner solar system from \textit{Helios},
\textit{Galileo}, \textit{Ulysses} and \textit{Pioneers} 8/9 show a
sharp drop-off in the IPD density beyond $1 \,$AU, and confinement of
the dust particles within $30^{\circ}$ of the ecliptic
plane\cite{Grun2001}.  It is difficult to formulate a mechanism that
produces a long-lived population of dust out of the ecliptic
plane\cite{Nesvorny2011}, so the bulk of the IPD material is thought
to reside at low inclination angles with respect to this plane and
models of the dust distribution support this\cite{Poppe2016}.
Interestingly, \textit{Ulysses} measurements far above the ecliptic
found a continuum level of particle events associated with the planar
inflow of interstellar dust from the local hot bubble\cite{Grun2001}.
These dust particles are very small, with characteristic radii
$1{-}10 \,$nm, so do not effectively reflect sunlight at optical
wavelengths.  As a result, there is no expectation that IPD light is sourced
far from the plane of the ecliptic.

In the outer Solar system, few \textit{in situ} dust measurements
exist.  \textit{Pioneers} 10 and 11 carried detectors that measured
the flux of $5 {-} 10 \, \mu$m particles\cite{Humes1980}.
\textit{Pioneer 10} reported data to $18 \,$AU\cite{Landgraf2002}.
Pioneer 11 made continuous measurements to $\sim 9 \,$AU and crossed
the $3.7 {-} 5 \,$AU region three times (once outbound and twice while
transiting from Jupiter to Saturn), finding consistent results each
time\cite{Dikarev2002}.  More recently, the Student Dust Counter (SDC)
on New Horizons has measured the flux of $0.5 {-} 5 \, \mu$m dust
grains from $5$ to $30 \,$AU\cite{Poppe2010, Szalay2013, Poppe2016}.
These measurements suggest an order of magnitude drop of the dust flux
from $1$ to $5 \,$AU, followed by a flattening of the particle flux to
at least $20 \,$AU.  Recently, a model has been generated that is
consistent with all of the in situ measurements\cite{Poppe2016}; the
predictions of this model scaled to a quantity that should follow the
IPD light intensity are shown in Supplementary Fig.~7.  From these
\textit{in situ} measurements, we would infer that the IPD population
in the region over which the observations are performed is small and
decreasing with distance.

In addition to \textit{in situ} measurements, Pioneer 10 and 11
observed the optical-wavelength intensity of the background through
the Solar system with a two-band imaging
photopolarimeter\cite{Pellicori1973}.  The Pioneer 10 measurements
from $2.4$ to $3.2 \,$AU exhibit a factor of $>25$ decrease in the
surface brightness of two survey regions\cite{Hanner1974}, both
measured at $\theta_{\odot} > 102^{\circ}$.  The measurements beyond
$3.25 \,$AU are individually consistent with zero; averaging over
these four measurements gives a $2 \sigma$ upper limit on the IPD
brightness of $< 4.9 \,$\nw, using the known
conversion\cite{Leinert1998} between $S_{10}(V)$ and \nw.  The Pioneer
10 measurements are shown in Supplementary Fig.~7,
and the upper limit on $\lambda I_{\lambda}^{\rm IPD}$ is listed in
Supplementary Table 2.  No published analyses of IPD light in the
Pioneer 11 data are available.

In the full set of 255 images, the LORRI data show no surface
brightness change with heliocentric distance consistent with a
variable contribution from the IPD, nor
with viewing angle through the plane of the ecliptic.  

\paragraph{Residual Starlight.}
There are two contributions to $\lambda I_{\lambda}^{\rm RS}$ in these
data: (\textit{i}) that from the unmasked wings of the PSF, and
(\textit{ii}) that from sources below the masking threshold.  To
calculate the residual starlight from the unmasked wings, we use the
USNO-B1 catalog and measured LORRI PSF to simulate each flight image.  These
images are then masked with the flight mask, and the mean of the
unmasked pixels is computed.  We estimate that the residual starlight
is negligible compared to the mean sky brightness in these images (see
Supplementary Table 2).

The residual starlight from stars with $R_{\rm L} > 17.7$ is
challenging to calculate from real catalogs as they do not approach
the required depth of $R \sim 25$.  As a result, we use the {\sc
  trilegal} model to estimate the faint star flux, which models star
fields as a function of position on the sky, photometric system,
assumed stellar IMF, binary fraction, the sun's position, and various
parameters describing the Milky Way's thin disc, thick disc, halo, and
bulge\cite{Girardi2005}.  The model returns catalogs of stars
consistent with the observed number counts and known populations of
stars.  The number counts are returned to high precision, and {\sc
  trilegal} performs particularly well away from the galactic plane
where the galaxy model is relatively simple.  For each field's
position, we generate ten {\sc trilegal} simulations of a
$0^{\circ}.3 \times 0^{\circ}.3$ field corresponding to the LORRI
image, complete to $R=32$.  For each simulation, we compute the mean
surface brightness of the corresponding image.  This results in the
$\lambda I_{\lambda}^{\rm RS}$ from faint sources listed in
Supplementary Table 2.  Even at the relatively faint masking threshold
we apply, the residual flux from faint stars contributes a surface
brightness comparable to the expected COB.

\paragraph{Diffuse Galactic Light.}
At optical wavelengths, dust in the galaxy reflects the local
interstellar radiation field, and may also luminesce\cite{Ienaka2013}.
Similarly to the ecliptic dependence of light from the IPD, the DGL is
brightest in the galactic plane and relatively faint at high galactic
latitudes.  Early Pioneer 10 measurements\cite{Toller1981} found a factor
$> 10$ reduction between the DGL measured on the galactic plane and at
the poles, and suggest a brightness of $\sim 150 \,$\nw\ and
$\sim 10 \,$\nw, respectively.  The implication is that nowhere on the
sky can we ignore the contribution from the DGL.  Since, on small
scales, the spatial variation of the DGL \cite{Zemcov2014} is
fractions of a \nw\, in these LORRI data the primary effect is that of
an overall surface brightness in the images.  Due to LORRI's broad
optical passband and the limited number of observed fields, with the
LORRI data alone the DGL would be impossible to disentangle from the
COB.

The dust grains responsible for the DGL are also heated by the
interstellar radiation field (ISRF) and emit this energy thermally in
the far-IR.  As a result, the DGL is highly correlated with
$100 \, \mu$m emission in the optically thin limit where the optical
photon scattering is simple\cite{Ienaka2013}.  Here we take advantage
of this correlation and the excellent $100 \, \mu$m all-sky surface
brightness maps available\cite{Schlegel1998, Miville2005} to estimate
the contribution of the DGL to the optical surface brightness in each
of the four fields.  We have restricted our attention to high galactic
latitude fields with $I_{100 \mu {\rm m}} < 10 \,$MJy sr$^{-1}$ in
order to avoid optically thick dust as part of the data cut process.
This allows us to take advantage of the linear relationship between
thermal emission intensity and optical surface brightness.

We estimate the absolute surface brightness of the DGL in each field
via the following relation:
\begin{equation}
\lambda I_{\lambda}^{\rm DGL}(\lambda,\ell,b) = \nu \langle
I_{\nu}(100 \mu \mathrm{m}) \rangle \cdot \bar{c}_{\lambda} \cdot
d(b), 
\end{equation}
where $\nu \langle
I_{\nu}(100 \mu \mathrm{m}) \rangle$ is the $100 \, \mu$m surface brightness
averaged over the field, $\bar{c}_{\lambda}$ is the conversion from
thermal emission intensity to optical surface brightness formulated
below, and $d(b)$ is a function that accounts for the change in
$c_{\lambda}$ due to scattering effects as a function of galactic
latitude.

To estimate $\nu \langle I_{\nu}(100 \mu \mathrm{m}) \rangle$ we
compute the Improved Reprocessing of the IRAS Survey (IRIS)
$100 \, \mu$m image\cite{Miville2005} for each pixel in the LORRI
images.  We then subtract $0.8 \,$MJy sr$^{-1}$ to account for the CIB
brightness at $100 \, \mu$m\cite{Puget1996,Fixsen1998}, yielding the
brightness of the dust in each field.  

There are a variety of measurements of the scaling between the surface
brightness in the optical/near-IR and at $100 \, \mu$m,
$c_{\lambda} = \lambda I_{\lambda} ({\rm opt}) / \nu I_{\nu} (100 \,
\mu {\rm m})$
(which is sometimes quoted as
$b_{\lambda} = I_{\nu}({\rm opt}) / I_{\nu} (100 \, \mu {\rm m})$).
We estimate $b_{\lambda}$ by fitting the mean `ZDA04'
model\cite{Zubko2004} to a compilation of measurements of
$b_{\lambda}$\cite{Ienaka2013}, which yields a best-fitting
$c_{\lambda}$ and its uncertainty through multiplication by a factor
of $ 10^{-6} \cdot (100 \, \mu \mathrm{m} / 0.655 \, \mu \mathrm{m})$.
We then compute the wavelength-averaged value $\bar{c}_{\lambda}$ by
integrating $c_{\lambda}$ weighted by the LORRI bandpass.  This gives
$\bar{c}_{655 {\rm nm}} = 0.49 \pm 0.13$.

Finally, we estimate $d(b)$ using the relation:
\begin{equation}
d(b) = d_{0} (1 - 1.1 g \sqrt{\sin(|b|)})
\end{equation}
where $d_{0}$ is a normalizing factor and the Henyey-Greenstein
parameter $g$ is the asymmetry factor of the scattering phase
function\cite{Jura1979}.  To estimate $g$, we compute the
bandpass-weighted mean of an observation-constrained model for the
high-latitude diffuse dust component of the DGL\cite{Draine2003}, and
take the allowed variation in the mean from measurement and modeling
errors as its uncertainty, yielding $g=0.61 \pm 0.10$.  To determine
$d_{0}$, we normalize $d(b)$ at $b=25^{\circ}$ to be consistent with
previous measurements\cite{Lillie1976}.  As a check, we investigate
the effect of using a larger estimate for $g$ consistent with
scattering from dense molecular clouds below.

From this set of information, $\lambda I_{\lambda}^{\rm DGL}$ can be
computed.  The value of $\lambda I_{\lambda}^{\rm DGL}$ in each field
is listed in Supplementary Table 2.  Due to the relatively high
galactic latitude and small size of the LORRI fields, and the large
effective smoothing in the IRIS maps, we find it is unnecessary to
account for the spatial variation in DGL in these fields.

As a check of the relatively uncertain direct DGL subtraction, we also
compute a linear fit of the diffuse sky brightness minus the residual
starlight against the $100 \, \mu$m surface brightness in each of the
four fields via:
\begin{equation}
\lambda I_{\lambda}^{\rm diffuse} - \lambda I_{\lambda}^{\rm RS} =
\bar{c}_{\lambda}^{\prime} \cdot \nu \langle I_{\nu}(100 \, \mu {\rm m}) \rangle + \lambda I_{\lambda}^{\rm
  resid} 
\end{equation}
where $\bar{c}_{\lambda}^{\prime}$ and $\lambda I_{\lambda}^{\rm resid}$ are the
parameters of the fit.  We find a best-fitting
$\bar{c}_{\lambda}^{\prime} = 0.40 \pm 0.27$ and COB results consistent with
the values determined via the direct subtraction method, but with much
larger uncertainties since, in this fitting method, errors in the other
DGL model parameters are folded into the measurement.  When systematic
errors are included, the estimates from both methods are similar in
both absolute value and uncertainty, as expected.

\paragraph{Extinction correction.}  After accounting for the
astrophysical foregrounds in the LORRI images, we retrieve the
residual sky brightness measurements shown in Supplementary Fig.~8
and listed as $\lambda I_{\lambda}^{\rm resid}$ in Supplementary Table
2.  The per-field measurements are computed as the weighted mean of
the individual exposures, where the weights are the inverse error on
the mean in each $10 \,$s exposure calculated from the variance of
unmasked pixels.  We take the uncertainty on the mean to be the the
standard deviation of the individual exposures in each field.

To propagate these field averages to a measurement of the COB, it is
necessary to correct for the effect of galactic extinction.  One of
the field averages is negative, for which an extinction correction is
unphysical, so we choose to compute the mean residual brightness and
then apply an equivalently-generated extinction correction to that
quantity.  We first compute the uncertainty-weighted mean of the
fields, and find $\lambda I_{\lambda}^{\rm resid} = 4.5 \pm 6.9 \,$\nw.

Next, we estimate the extinction correction by computing the mean of
the two $A_{R}$ predicted by two
models\cite{Schlafly2011,Schlegel1998} in each field.  We then compute
the mean of the four field extinction measurements weighted by the
same uncertainty weights used in the mean intensity computation.  This
yields an extinction correction of $A_{R} = 0.11$ mag.  However, galactic
extinction comprises two components, namely scattering and absorption.
For point source observations, both of these remove light out of the
line of sight, but since the COB is isotropic, the scattered light is
replaced by light from other lines of sight in a conservative
fashion.  The proportion of these effects is roughly 40 \% absorption
and 60 \% scattering, so our actual extinction correction is
$A_{R} = 0.05$ mag.  Applying this to $\lambda I_{\lambda}^{\rm resid}$,
we find $\lambda I_{\lambda}^{\rm COB} = 4.7 \pm 7.3 \,$\nw, where the
errors are purely statistical.  Extinction corrections do not apply to
the systematic errors as they are all due to local mechanisms.

\paragraph{Systematic and Foreground Error Estimation.}
The errors in this measurement can broadly be categorized as:
statistical; systematics in the instrument, where we include the
calibration uncertainty in this category; and systematics in the
astrophysical foreground accountancy.  The uncertainties in the
measurement are summarized in Supplementary Table 3 and Figure 3, and
their derivation is described in detail in this section.  Mean values
are always computed using the same field-to-field statistical weights
as used in the average COB brightness calculation.

The statistical uncertainties are computed from the variance of a
number of independent measurements, and as a result fold in a variety
of sources of noise (detector, read out, photon, \textit{etc.}) in an
indistinguishable way.  Based on a cursory inspection of the data and
the known photocurrent, the noise is dominated by bit noise from the
analog to digital converter, which suggests that in future
measurements increased integration times would be beneficial.

As there are several steps in the data reduction, there are a
corresponding number of potential errors in the instrumental
corrections and data analysis we apply.  First, the reference frame
subtraction would have an error associated with it.  However, because
we later subtract the reference pixel values, we are removing a large
part of the frame to frame variation that would lead to some offset
error in this step.  As a result, we fold all of the uncertainty
associated with reference value/dark current into that step's error,
as described below.  Second, the application of a inter-pixel gain
correction may have some intrinsic error, but because we perform
photometric calibration after the flat field correction and we are not
interested in the spatial structure of the images, errors in the
applied pixel-to-pixel response should have a negligible effect on the
final result.  The only situation in which such an error could have a
measurable impact is if the regions immediately surrounding the
calibration stars had some local inter-pixel response different from
the bulk of the detector array, and different than that measured
during laboratory testing.  To guard against this, we used calibration
stars in random positions on the detector array and in multiple
fields, and have shown the calibration to be consistent through the
observation period.  The photometric calibration uncertainty captures
the remaining uncertainty from this effect.

As they are electrically identical to the photo-responsive pixels, and
share the same read out chain, we have no expectation of or evidence
for the reference pixel subtraction leading to any misestimation of
the overall array offset.  Extra variance in the final image
brightness from the intrinsic measurement error of the $256$ reference
pixels that varies frame to frame is naturally accounted in the
statistical uncertainty estimate.  However, as the reference pixels
are shaded from incident photons by a metal slat, it is possible that
there is a $< 20\,$\% reduction in the dark current due to
electromagnetic coupling to the shade (A.~Reinheimer, private
communication).  To estimate the uncertainty from this effect, we
compute $20 \,$\% of the expected dark current in the pixels, which
would cause a spurious image offset of $0.9 \,$\nw.  Because this
would be in the direction of an under-subtraction, this error is in the
positive-going direction and should be applied uniformly through the
observations.

Optical ghosts were identified to be present in the central $r < 50$
pixel region of the LORRI images during laboratory
testing\cite{Cheng2008}.  The position and brightness of these ghosts
depend on bright stars slightly off the imaged field, and they may be
fainter than can be easily detected in the images.  Though we masked
these ghosts from our science images, ghosts fainter than the surface
brightness limit to which we masked may be present.  As a check of our
masking procedure, we mask the full $r < 50$ region from the science
images and recompute the resulting COB brightness through the entire
analysis pipeline.  When this augmented mask is applied to the images,
we find a modest change to $\lambda I_{\lambda}^{\rm inst}$ of
$-0.1 \,$\nw, \textit{i.e.} in the sense of an increased
$\lambda I_{\lambda}^{\rm resid}$.  Unmasked optical ghosts would have
the opposite sign, so we conclude there is no evidence for excess
surface brightness from optical ghosts in these data.

Based on the dispersion of the aperture photometry measurements, we
estimate the raw photometric calibration uncertainty of this
measurement to be $\pm 7.3 \,$\%.  This compares well with the
$0.25 \,$mag catalog standard deviation quoted as the per source
photometric accuracy of the USNO-B1 catalog, which for 8 objects would
give a photometric accuracy of $7.3 \,$\%.  USNO-B1 was ultimately
calibrated from the Tycho-2 catalog, which itself has been calibrated
to an accuracy of $\pm 2 \,$\%\cite{MA2005}.  We therefore estimate
the absolute photometric uncertainty of this study to be $\pm 8 \,$\%
of $\lambda I_{\lambda}^{\rm diffuse}$, which corresponds to $\pm 3.8
\,$\nw\ on $\lambda I_{\lambda}^{\rm COB}$.

The uncertainty in the solid angle of the beam also plays a role in
the ultimate calibration uncertainty of this study.  Based on
half-half PSF stacking jack knife tests, we estimate the solid angle
of the LORRI beam is known to $\pm 4 \, \%$.  This propagates to a
$4 \,$\% uncertainty on $\lambda I_{\lambda}^{\rm diffuse}$, which
corresponds to $\pm 1.9 \,$\nw\ on $\lambda I_{\lambda}^{\rm COB}$.
This uncertainty also includes the error in $\lambda
I_{\lambda}^{\ast} + \lambda I_{\lambda}^{\rm RS}$ due to our
imperfect knowledge of $\Omega_{\rm beam}$ on the conversion from flux
to surface brightness.

The astrophysical foregrounds present in this study all have errors in
their estimation.  We have argued for a low level of IPD light in the
outer Solar system, but explicitly quote the full $1 \sigma$
uncertainty on the $R_{\odot} > 3.3 \,$AU Pioneer 10
measurements as an upper limit on the total IPD light contribution.

To account for the effect of the USNO-B1 photometric calibration
uncertainty, we compute the estimate for
$\lambda I_{\lambda}^{\rm diffuse}$ after randomizing the reported
magnitude of each source from the catalog by $\pm 0.25 \,$mag using a
random Gaussian deviation per source.  This has the effect of
modifying the mask radius to be either inappropriately small or large,
depending on the sign of the randomization.  Over many masked sources,
this should adequately probe the photometric error from the catalog
calibration.  Based on this calculation, we estimate the error to be
$\pm 0.1 \,$\nw on 
$\lambda I_{\lambda}^{\rm RS}$, resulting in an error on the COB
of $\pm 0.1\,$\nw.

To calculate the variation in $\lambda I_{\lambda}^{\rm RS}$ due to
sample variance of the faint stars in our fields, we calculate the
variance of ten realizations of the {\sc trilegal} model for each field,
complete to $R=32$.  As these are relatively high galactic latitude
fields, we find the standard deviation in
$\lambda I_{\lambda}^{\rm RS}$ is $0.6 \,$\nw\ over the set,
corresponding to $\pm 0.7 \,$\nw\ on $\lambda I_{\lambda}^{\rm COB}$.

We compute the error in our estimate for the DGL brightness by
propagating the error on the parameters in the various input
functions.  Specifically, we use
$\nu I_{\nu}(100 \, \mu \mathrm{m}) = 0.80 \pm
0.25$\cite{Puget1996,Fixsen1998},
$g=0.61 \pm 0.10$\cite{Sano2016}, and
$\bar{c}_{\lambda} = 0.49 \pm 0.13$\cite{Ienaka2013} as being
consistent with the existing measurements.  Propagating these errors
through the DGL model gives
$\sigma(\lambda I_{\lambda}^{\rm DGL}) = \{-8.7,+8.2\} \,$\nw,
resulting in an uncertainty of $\{-9.1,+8.6\} \,$\nw\ on
$\lambda I_{\lambda}^{\rm COB}$.  As a check of the effect of the
relatively uncertain value of $g$, we use a compilation of results
based on measurements of dense molecular clouds\cite{Sano2016} that
have a different mean scattering asymmetry factor $g=0.75 \pm 0.1$ and
find $\lambda I_{\lambda}^{\rm COB} = 3.8 \,$\nw, well within the
quoted systematic uncertainty.

Finally, the galactic extinction correction is based on models that
themselves have uncertainties.  To bracket these, we take the allowed
difference between the two extinction
models\cite{Schlafly2011,Schlegel1998} as our best estimate for
$\sigma_{A_{\rm R}}$.  Over the four science fields, we compute the
uncertainty-weighted allowable variation in the extinction correction
to be a factor of $0.01$ in surface brightness, which corresponds to a
negligible error in $\lambda I_{\lambda}^{\rm COB}$.

\paragraph{Data Availability.}  The data that support the findings
of this study are available from the NASA Planetary Data System at
\url{http://pds-smallbodies.astro.umd.edu/data_sb/}\\
\url{missions/newhorizons/index.shtml}.

\noindent \textbf{References}
\bibliography{nh_lorri}

\section*{Acknowledgements}
\vspace*{-15pt} We would like to thank both the New Horizons
science team and the LORRI instrument team for their decades of
dedicated effort designing, building and flying such a complex
mission.  Many thanks to H.~Weaver and M.~Richmond for useful
discussions during the course of this work, and to the referees whose
depth of knowledge and insights significantly improved the work.  The
New Horizons launch, Jupiter fly-by, and cruise phase data
sets were obtained from the Planetary Data System (PDS).
A.~R.~P.~gratefully acknowledges NASA Planetary Atmospheres grant
\#NNX13AG55G.

\section*{Author contributions}
\vspace*{-15pt} M.~Z.~developed the analysis and systematic error
assessment pipeline, performed the foreground analysis, and wrote the
first draft of the paper. P.~I.~generated the data cuts and determined
the photometric calibration of the instrument.  C.~N.~collected the
data from the archive and performed a variety of data quality checks.
A.~C., C.~L.~and A.~P.~worked on various aspects of foreground
analysis and provided input on the low level analysis and workings of
the instrument.  All coauthors provided feedback and comments on the
paper.

\section*{Additional information}
\vspace*{-15pt}
\noindent \textbf{Supplementary Information} accompanies this paper at \\
\url{http://www.nature.com/naturecommunications}.

\noindent \textbf{Competing financial interests:} The authors declare no
competing financial interest. Reprints and permission information is
available online at \\
\url{http://npg.nature.com/reprintsandpermissions/}.

\noindent \textbf{How to cite this article:} Zemcov, M.~\textit{et
al.} Measurement of the Cosmic Optical Background using the Long Range
 Reconnaissance Imager on New Horizons.  Nat.~Commun. (2015).

\noindent This work is licensed under a Creative Commons Attribution 4.0
International License. The images or other third party material in
this article are included in the article’s Creative Commons license,
unless indicated otherwise in the credit line; if the material is not
included under the Creative Commons license, users will need to obtain
permission from the license holder to reproduce the material. To view
a copy of this license, visit
\url{http://creativecommons.org/licenses/by/4.0/}

\newpage

\setcounter{figure}{0}
\setcounter{table}{0}

\begin{center}
\textbf{{\Large Supplementary Information}}
\end{center}

\renewcommand{\figurename}{Supplementary Figure}
\renewcommand\thefigure{\arabic{figure}} 
\renewcommand{\tablename}{Supplementary Table}
\renewcommand\thetable{\arabic{table}} 

\begin{figure*}[h]
\centering
\includegraphics*[width=6.5in]{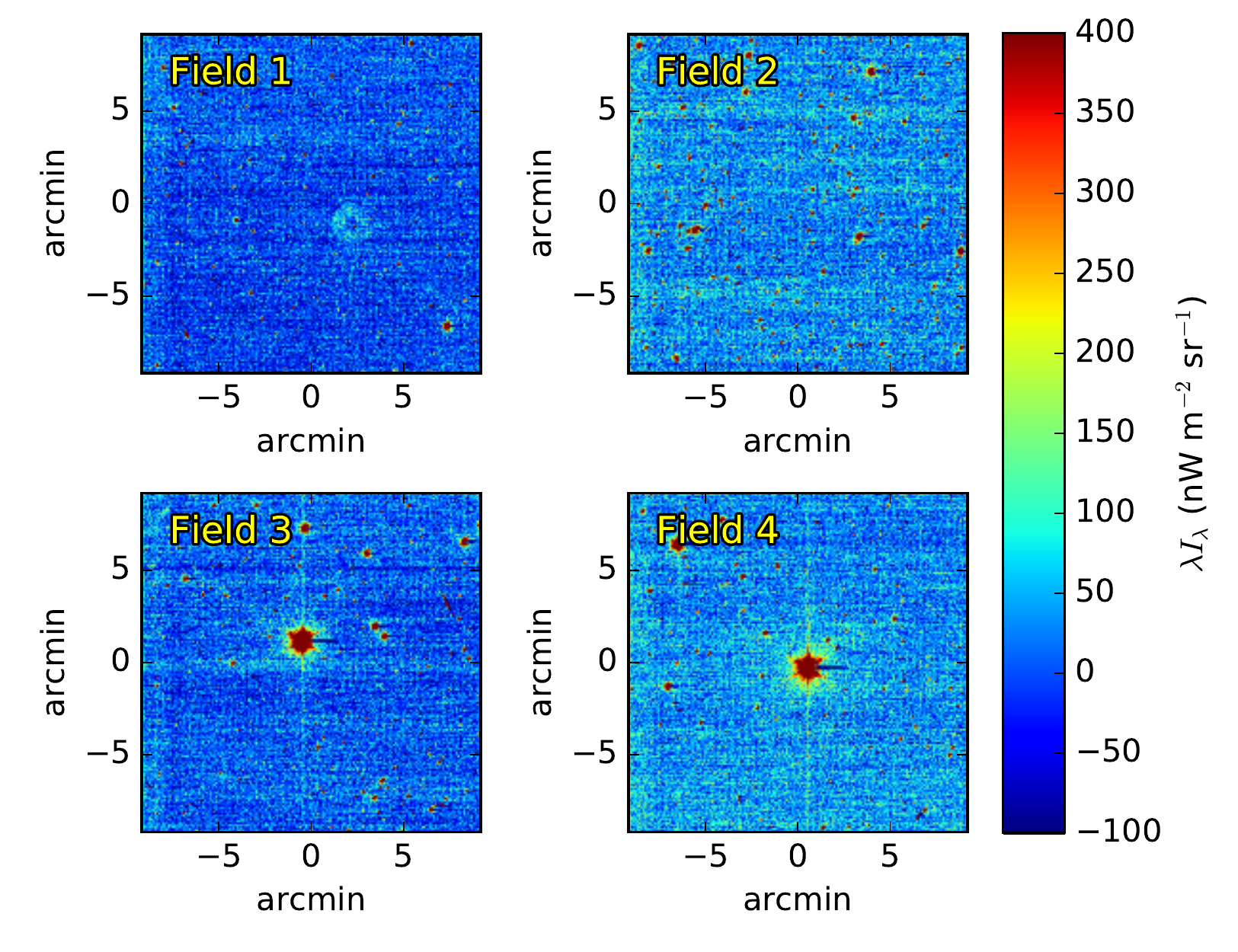}
\caption{\textbf{Reduced images of the four science fields used in this
  investigation, calibrated to surface brightness units.}  Each panel
  shows one $10 \,$s integration for each of the fields taken in
  rebinned $256 \times 256$ mode.  Stars and Neptune (the large source
  in fields 3 and 4) are clearly visible.  Field 1 exhibits optical
  ghosting from reflections in the field-flattening lens group.  All
  of these structures are masked in later processing step to allow us
  to calculate the mean surface brightness away from known
  sources.  \label{fig:images} }
\end{figure*}

\begin{figure*}[p]
\centering
\includegraphics*[width=6.5in]{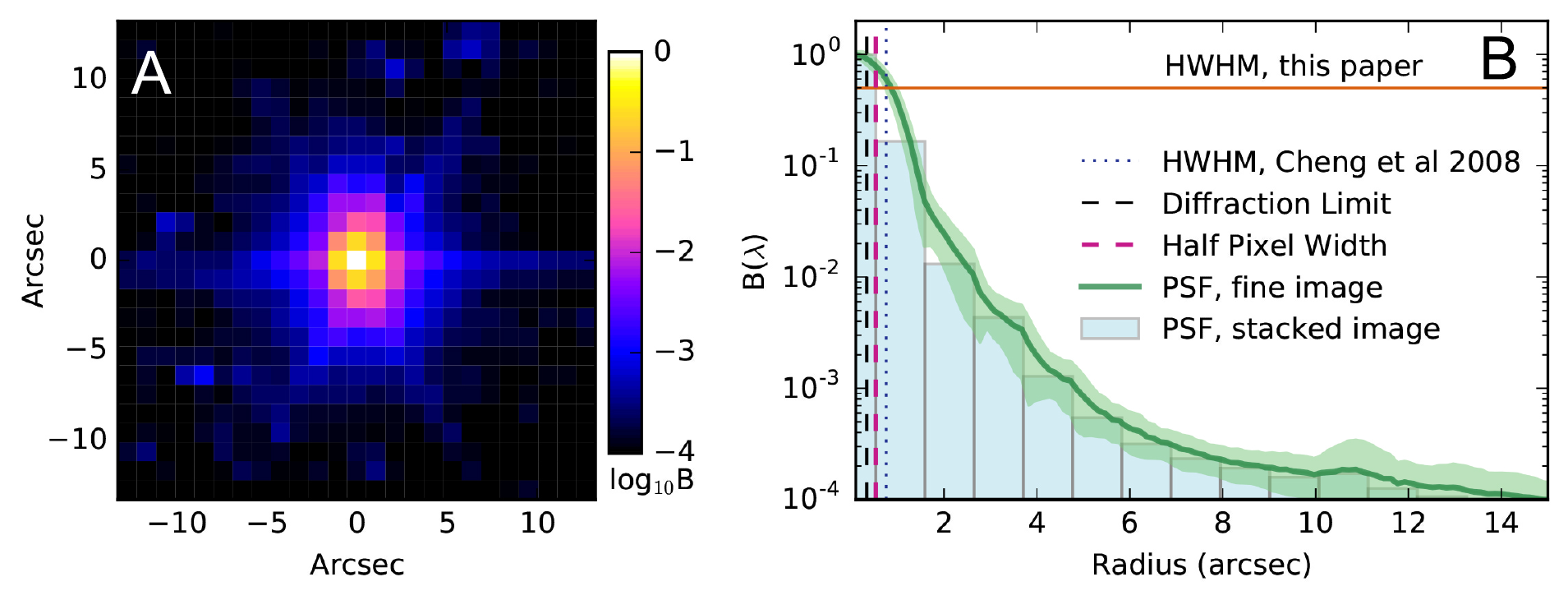}
\caption{\textbf{The point spread function (PSF) used for source masking in
  this work.} The 2-D PSF shown in panel A is produced by stacking
  star images on a fine pixel grid, and then averaging back into
  larger pixels\cite{Bock2013}.  Panel B is the annular average of
  the stacked PSF in the left panel (light blue columns), in
  comparison to the PSF of the original, fine stacked image (bold
  solid green line). The shaded region shows the standard deviation of
  the fine PSF in annuli.  Also shown are the diffraction limit for
  the LORRI telescope (thin dashed line), the half pixel width (thick
  dashed line), the half width at half maximum (HWHM) of our PSF
  (marked by the crossing point of the horizontal solid line and the
  fine PSF), and the HWHM from previous
  determinations\cite{Morgan2005, Cheng2008, Owen2012}.  The HWHM of
  our PSF measurement is consistent with the previous determinations.
  Any asymmetry of the PSF at low levels is likely due to a small
  pointing drift, but does not affect our analysis.  \label{fig:psf} }
\end{figure*}

\begin{figure*}[p]
\centering
\includegraphics*[width=6.5in]{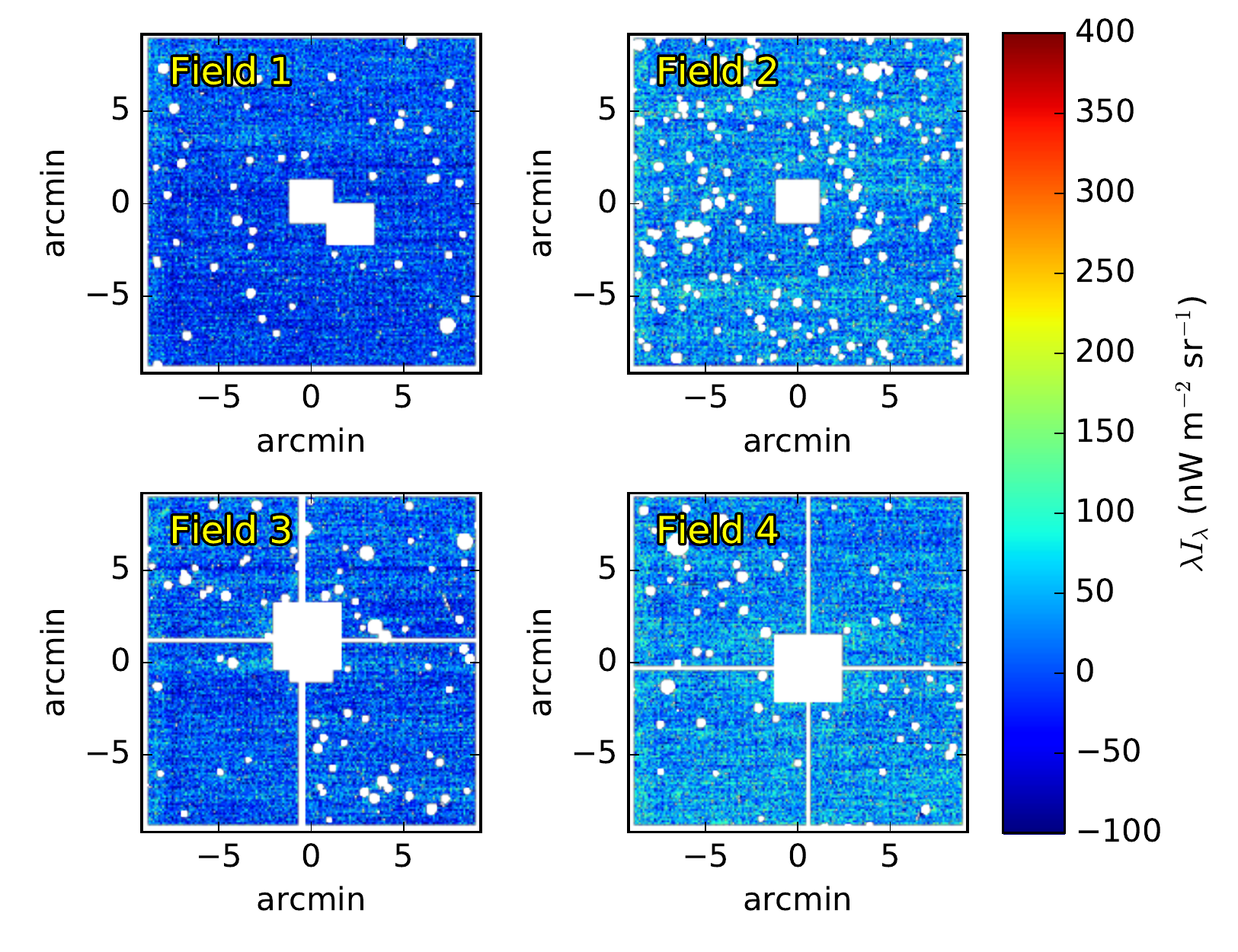}
\caption{\textbf{Masked, reduced images of the four science fields used in
  this investigation, calibrated to surface brightness.}  Each panel
  shows one $10 \,$s integration for each of the fields taken in
  rebinned $256 \times 256$ mode with the source and array masks
  applied.  The mask effectively removes images of stars, planets, and
  optical and electronic pathologies in the images.  The diffuse sky
  brightness $\lambda I_{\lambda}$ is computed from the mean of the
  unmasked pixels in these images.  The mild horizontal striping is
  due to noise correlations in the output amplifier circuit, and are
  naturally present in any image of a sequentially read Si detector
  close to the noise floor.  This effect causes the pixel variance to
  be a poor estimate of the absolute statistical error, though it
  remains a reasonable weight for averaging calculations where only
  changes in the exposure-to-exposure variance are
  relevant.  \label{fig:masked} }
\end{figure*}

\begin{figure*}[p]
\centering
\includegraphics*[width=6.5in]{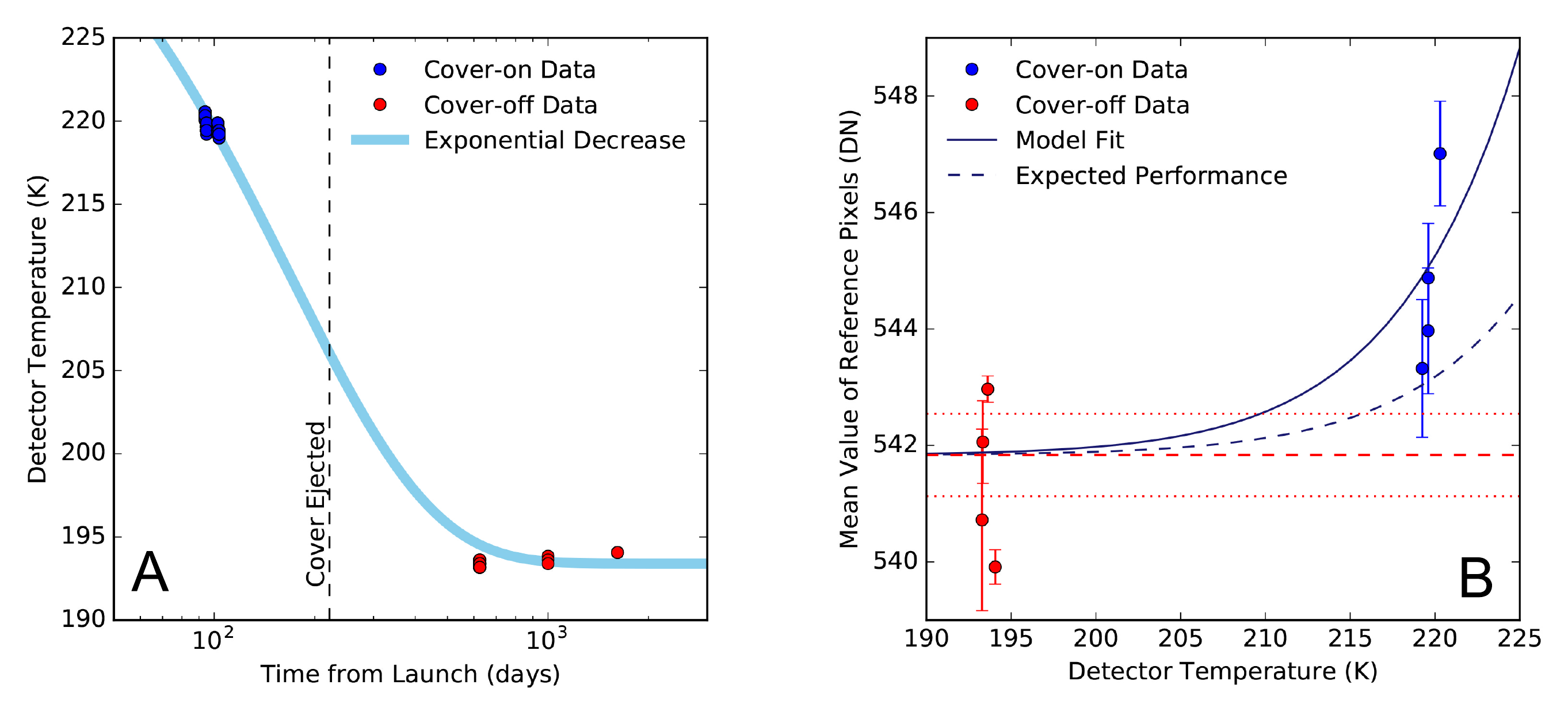}
\caption{\textbf{Temperature stability and reference pixel values over flight
  for both covered and uncovered data.}  Measurements of the LORRI CCD
  temperature versus time for both dust cover on (blue points) and
  dust cover off (red points) are shown in panel A.  We also indicate
  the time when the cover was ejected (dashed line) and the
  best-fitting exponential temperature decrease over the period of
  data collection.  The plotted uncertainties correspond to the
  standard deviation of the individual measurements.  Panel B shows
  the average value of the 256 dark reference pixels versus
  temperature over the flight.  We indicate the mean cover-off
  reference pixel value (red dashed lines) and $1 \sigma$ variation in
  the individual measurements (red dotted lines).  Finally, a model
  for the dark current expected in these devices from the
  manufacturers specification (dashed blue line) and a free-amplitude
  fit of the same model (solid blue line) are shown.  The dust-cover
  on data prefer an elevated dark current level above the baseline for
  this detector, but in either case the dark current is small at the
  $193 \,$K operating temperature of the science observations. The
  residual reference fluctuations can be explained by temperature
  fluctuations in the electronics chain and bias generation
  (H.~Weaver, private communication).  \label{fig:darkcurrent} }
\end{figure*}

\begin{figure*}[p]
\centering
\includegraphics*[width=6.5in]{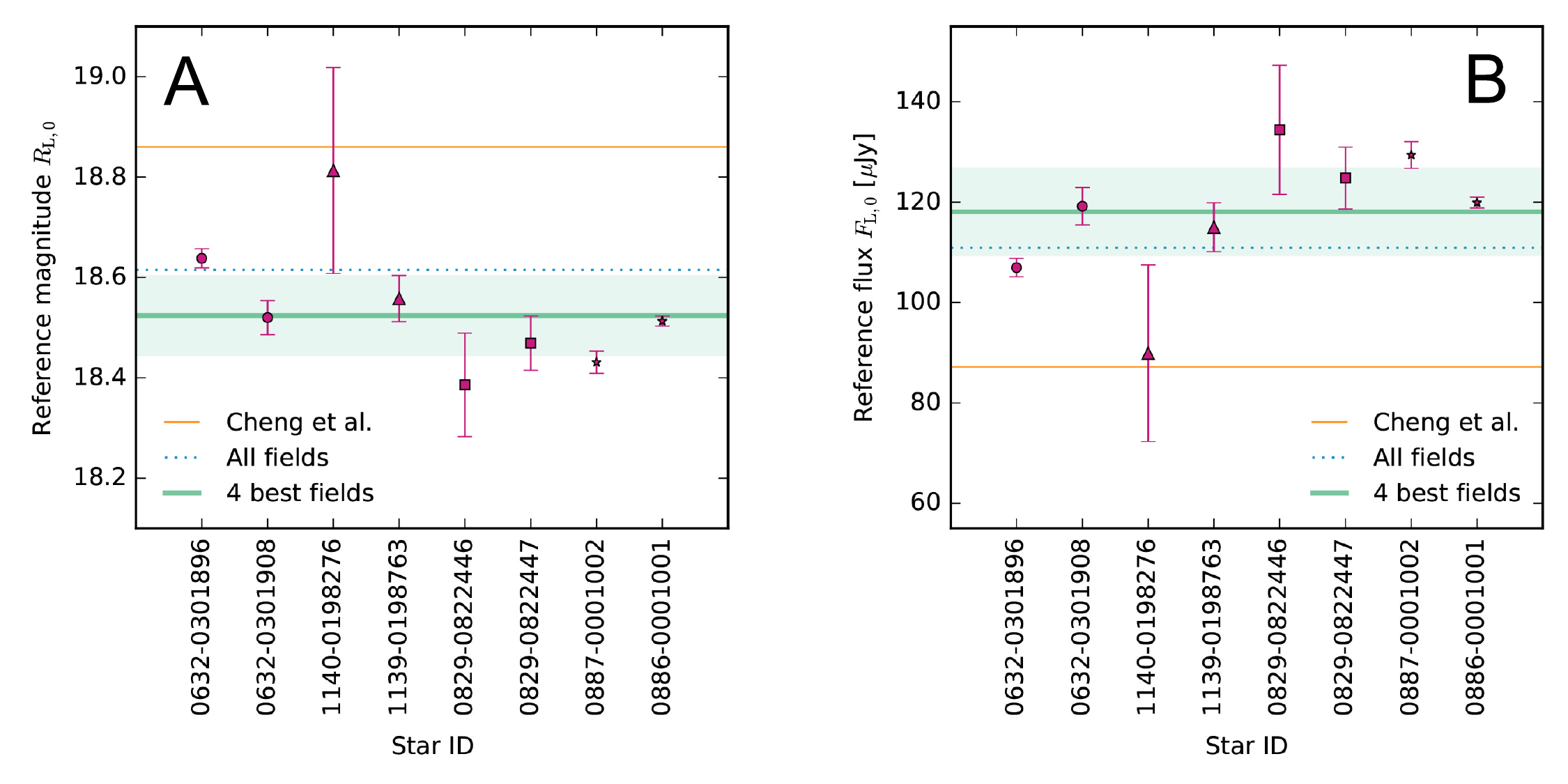}
\caption{\textbf{Summary of the photometric calibration in these LORRI data.}
  We plot the reference magnitude $R_{{\rm L},0}$ (panel A) and the
  reference flux $F_{{\rm L}, 0}$ (panel B). $R_{{\rm L},0}$ is
  calculated from aperture photometry as described in the text, and
  $F_{{\rm L},0}$ is calculated from $R_{{\rm L},0}$ using the PSF
  measurement.  The points are computed using multiple measurements of
  two different stars in each of the four fields (circle, field 1;
  triangle, field 2; square, field 3; star, field 4). The average
  magnitude and flux of the stars in the four science fields are shown
  with the $1 \sigma$ uncertainties, which we take as the best
  estimate of LORRI's photometric calibration.  We compute
  $R_{{\rm L},0}$ and $F_{{\rm L}, 0}$ from all the available
  cover-off data, which matches the science field data within the
  errors.  This shows that the reference magnitude and flux inferred
  from the four science fields are consistent with and representative
  of the entire sample.  We also show the preliminary LORRI
  photometric calibration\cite{Cheng2008} converted to
  $R_{{\rm L},0}$-band; though our calibration is inconsistent with
  this measurement, it is based on a single observation of a crowded
  field with high photocurrents and no measurement uncertainty was
  specified.  An improved photometric calibration of LORRI from the
  New Horizons team will be available in the near
  future.  \label{fig:calibration} }
\end{figure*}

\begin{figure*}[p]
\centering
\includegraphics*[width=5.87in]{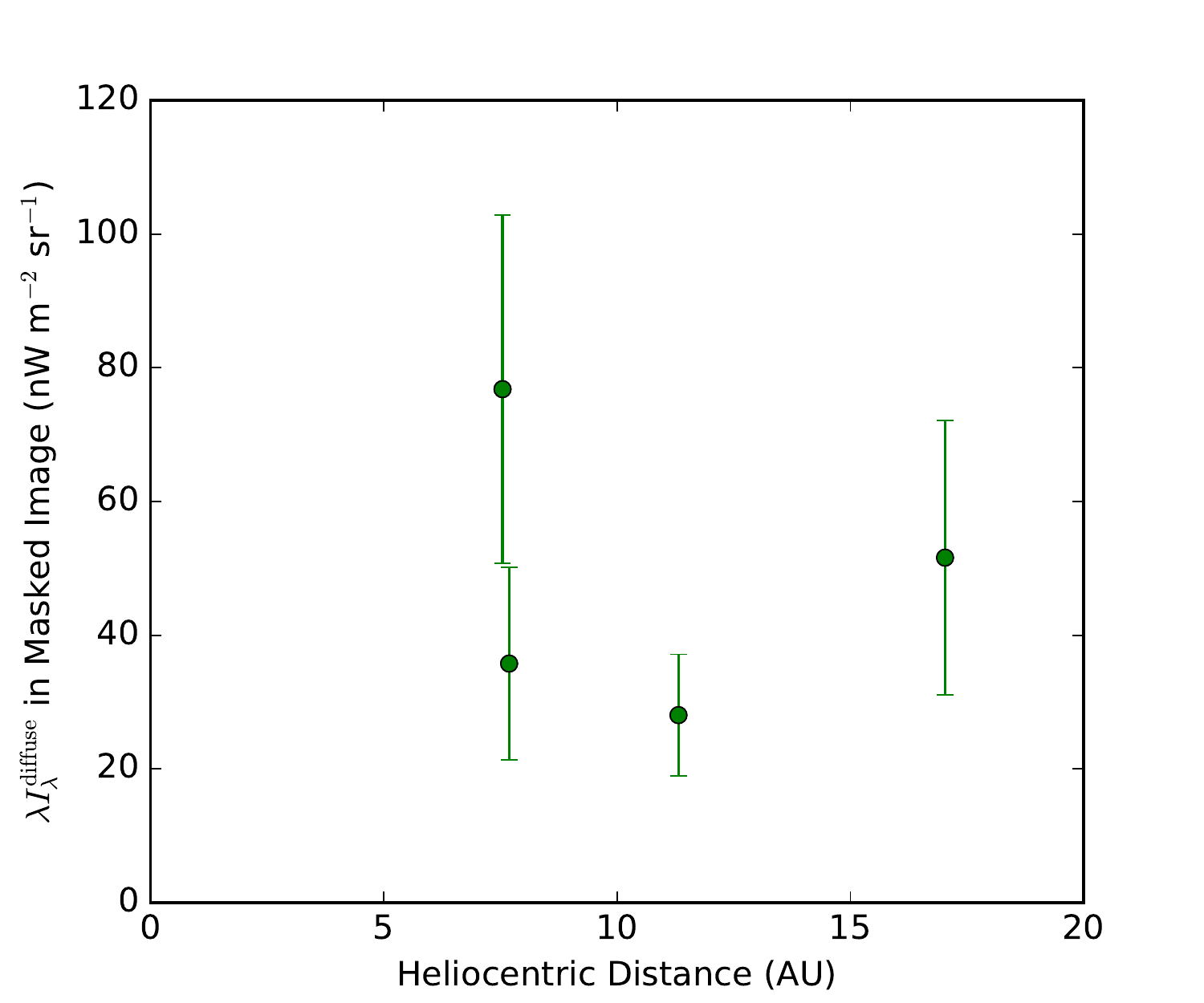}
\caption{\textbf{Measurement of $\lambda I_{\lambda}^{\rm diffuse}$ in the
  four science fields plotted as a function of heliocentric radius.}
  The points are computed from the variance-weighted mean of the
  individual $10 \,$s exposures in each field, and the plotted
  uncertainties correspond to the standard deviation of the mean pixel
  value of each exposure on a given field, which should provide a
  reasonable estimate of the statistical noise in the data.  These
  surface brightnesses still include emission from astrophysical
  foregrounds, which depend on field position in a complex way and
  must be estimated and removed to isolate the COB component.  The
  first two points have been offset slightly in the horizontal
  direction to improve clarity.  \label{fig:raw} }
\end{figure*}

\begin{figure*}[p]
\centering
\includegraphics*[width=6.5in]{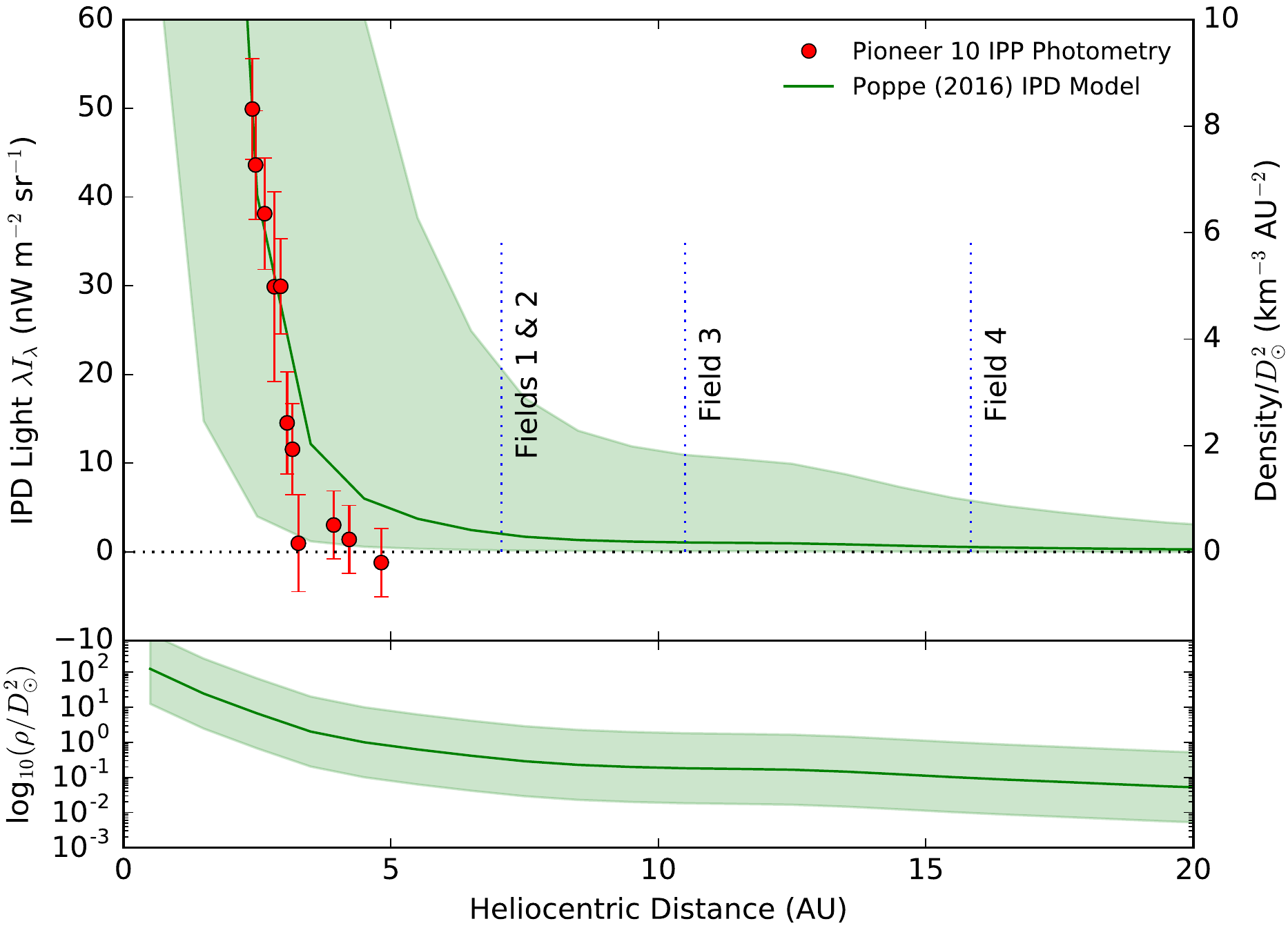}
\caption{\textbf{Surface brightness of the IPD and the $0.5 {-} 100 \, \mu$m
  particle density per $R_{\odot}^{2}$ as a function of heliocentric
  distance.}  We indicate the positions 
  of the LORRI measurements by the labeled blue dashed lines.  The red
  points show the photometric measurements from Pioneer 10
  converted to \nw\ and referenced to $R_{\rm L}$-band\cite{Hanner1974}.
  For comparison, the value of $\lambda I_{\lambda}^{\rm ZL}$ near the
  earth is $\sim 1000 \,$\nw\ at similar solar elongations, ecliptic
  latitudes, and wavelengths\cite{Leinert1998}.  These data indicate
  a significant drop from $2$ to $3.3 \,$AU, and are consistent with
  zero beyond the asteroid belt.  The green line shows the local IPD density
  predicted by the model of Poppe\cite{Poppe2016} divided by the
  square of the distance to the 
  sun to account for the diminishing of intensity of sunlight with distance.  The
  filled green region shows an estimate for the uncertainty in the
  density model.  Assuming that the IPD light is sourced by particles close to the
  observer, and that the light scattering ability of these particles
  is independent of $R_{\odot}$, this model should exhibit the same behavior as the IPD
  light intensity.  The model prediction follows the IPD light photometric
  measurements quite well.
  Although
  the absolute scaling between the dust flux and the IPD light is
  unknown, the general trend of a decrease from Earth followed by a
  flattening in the outer Solar system occurs in both measurements.
  Based on the IPD density model, we expect the IPD light to have a
  small and diminishing surface brightness in the plane of the outer solar
  system. 
  \label{fig:ipdlight} }
\end{figure*}

\begin{figure*}[p]
\centering
\includegraphics*[width=5.87in]{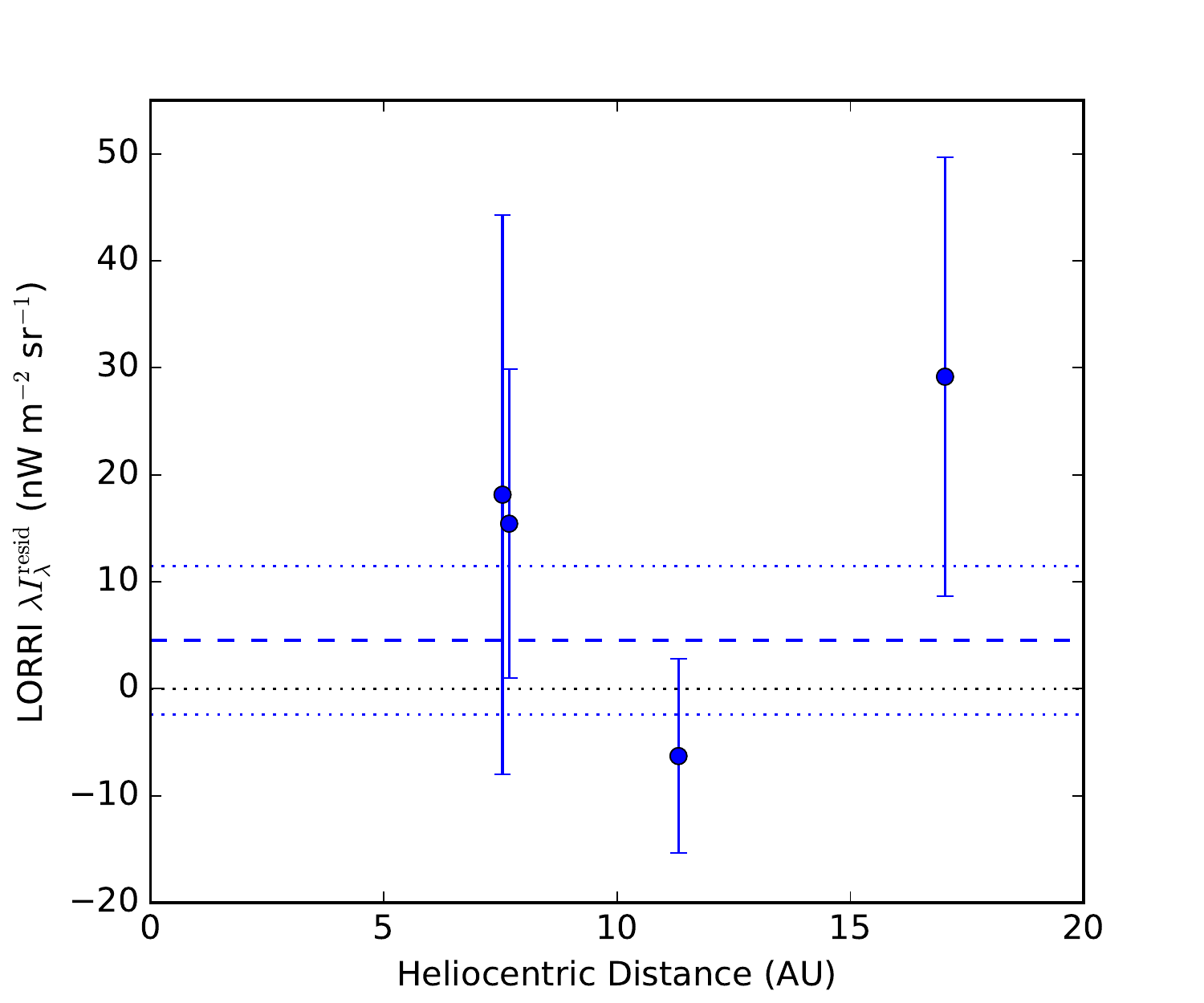}
\caption{\textbf{Residual surface brightness in the four LORRI
  science fields plotted as a function of heliocentric radius.}  The
  points are computed from the variance-weighted mean of the
  individual $10 \,$s foreground-subtracted exposures in each field,
  and the plotted uncertainties correspond to the standard deviation
  of the set of exposures, which should provide a reasonable estimate of the
  statistical noise in the data.  We also plot the
  uncertainty-weighted mean and error of the four measurements (blue
  dashed and dotted lines, respectively).  The unweighted mean lies at
  a larger value of 10.7 \nw\ in this plot.  The first two points have
  been offset slightly in the horizontal direction to improve clarity.  \label{fig:result} }
\end{figure*}

\begin{sidewaystable}[p]
\begin{center}
\caption{\textit{New Horizons} flight timeline. \label{tab:timeline}}
\begin{tabular}{lccccccc}
\hline
Date & Nominal Target & Number & Exposure Time &
$R_{\odot}$ & $\theta_{\odot}$ & 
$I_{100 \mu {\rm m}}$ & Field \\ 
 & & of Records & per Record (s) & (AU) & ($^{\circ}$) &  
(MJy sr$^{-1}$) & Number \\ \hline 

\multicolumn{8}{c}{\textit{Launch 2006 January 19}} \\
\hline

2006 April 23 & Dark & 120 & 10 & 1.8 & - & - & D1 \\

2006 April 24 & Dark & 59 & 10 & 1.8 & - & - & D2 \\

2006 May 2 & Dark & 120 & 10 & 1.9 & - & - & D3 \\

2006 May 3 & Dark & 60 & 10 & 1.9 & - & - & D4 \\

\hline
\multicolumn{8}{c}{\textit{Cover Ejection 2006 August 29}} \\
\hline

2006 August 31 & Messier 7 & 4 & 1 & 3.4 & 142.6 & 38.3 & R1 \\

2006 September 21 & $(\ell,b) = (13.1, 4.3)$ & 3 & 1 & 3.6 & 145.3 &
                                                                     91.5 & R2 \\

2006 September 22 & $(\ell,b) = (346.9, 31.5)$& 6 & 1 & 3.7 & 177.9 &
                                                                      7.6 & R3 \\

2006 September 24 & $(\ell,b) = (13.1, 4.3)$ & 3 & 1 & 3.7 & 145.5 &
                                                                     91.3 & R4 \\

2007 January 10 & Callirrhoe & 12 & 10 & 4.8 & 165.8 & 18.0 & R5 \\
 
\hline
\multicolumn{8}{c}{\textit{Jupiter Closest Approach 2007 February 28}} \\
\hline

2007 October 10 & Makemake & 10 & 10 & 7.6 & 85.5 & 2.4 & R6 \\

2007 October 10 & Haumea & 10 & 10 & 7.6 & 104.0 & 1.7 & 1 \\

2007 October 10 & Chariklo & 10 & 10 & 7.6 & 94.2 & 4.4 & 2 \\

2008 October 16 & Neptune & 3 & 10 & 11.3 & 104.8 & 3.6 & 3 \\

2010 June 23 & Neptune & 3 & 10 & 17.0 & 95.4 & 2.8 & 4 \\

2012 June 01 & Pluto & 5 & 10 & 23.3 & 167.8 & 147.7 & R7\\

2013 July 01 & Pluto & 90 & 10 & 26.7 & 166.6 & 148.0 &  R8 \\

2014 July 18 & Pluto & 48 & 10 & 29.9 & 165.7 & 147.0 & R9 \\

2014 July 20 & Pluto & 48 & 10 & 29.9 & 165.7 & 147.0 & R10 \\

\hline
\multicolumn{8}{c}{\textit{End of Public Cruise Records 2014 July 20}} \\

\hline
\end{tabular}
\end{center}
\end{sidewaystable}

\begin{table}
\centering
\caption{Foreground contributions to $\lambda I_{\lambda}^{\rm
    meas}$. \label{tab:foregrounds}}
\footnotesize
\begin{tabular}{l|ccccccc}
\hline
Field & $\lambda I_{\lambda}^{\ast}$ &
$\lambda I_{\lambda}^{\rm diffuse}$ &
$\lambda I_{\lambda}^{\rm IPD}$ & 
$\lambda I_{\lambda}^{\rm RS}$ from PSF wings & 
$\lambda I_{\lambda}^{\rm RS}$ from faint stars & 
$\lambda I_{\lambda}^{\rm DGL}$ & 
$\lambda I_{\lambda}^{\rm resid}$ \\ 

& \multicolumn{7}{c}{\textit{(All \nw)}} \\ \hline

1 & 674& 35.8 & $< 2.4, 1 \sigma$ & 0.02 & 12.7 & 7.7 & $15.4 \pm
                                                        14.4$ \\

2 & 967 & 76.8 & $< 2.4, 1 \sigma$ & 0.13 & 7.8 & 50.8 & $18.1 \pm
                                                         26.2$ \\

3 & 408 & 28.0 & $< 2.4, 1 \sigma$ & 0.01 & 6.7 & 27.6 & $-6.3 \pm 9.1$\\

4 & 242 & 51.6 & $< 2.4, 1 \sigma$ & 0.01 & 3.6 & 18.8 & $29.2 \pm 20.5$ \\

\hline
\end{tabular}
\end{table}

\begin{table}
\centering
\caption{Error budget for this measurement.  \label{tab:errors}}
\begin{tabularx}{\textwidth}{Xcc}
\hline
Error & Parameter Modification & Uncertainty in COB$^{\ast}$ \\
& & (\nw) \\ \hline

Statistical & $\lambda I_{\lambda}^{\rm COB} \pm 7.3 \,$\nw & $\pm
7.3$ \\[10pt] \hline 

Dark Current & $\lambda I_{\lambda}^{\rm inst}
+0.9 \,$\nw & $-0.9$\\[10pt]

Optical Ghosts & $\lambda I_{\lambda}^{\rm inst}
- 0.1 \,$\nw & $-$ \\[10pt]

Photometric Calibration Uncertainty & 8\% of $\lambda I_{\lambda}^{\rm
  diffuse}$ & $\pm 3.8$ \\[10pt]

$\Omega_{\rm beam}$ Uncertainty & 4\% of $\lambda I_{\lambda}^{\rm diffuse}$
& $\pm 1.9$ \\[10pt] 

Aperture Photometry Loss & $0.02$\% of calibration factor & $-$ \\ [10pt]
\hline 


Masking Bright Galaxies & $\lambda I_{\lambda}^{\rm diffuse} +0.006
                          \,$\nw & $-$ \\[10pt]

IPD Light Uncertainty & $\lambda I_{\lambda}^{\rm IPD} + 2.4 \,$\nw &
$-2.6$ \\[10pt]

$0.25 \,$mag USNO-B1 Photometry Uncertainty (per Source) on Source
  Mask & $\lambda 
I_{\lambda}^{\rm RS} \pm 0.1 \,$\nw & $\pm 0.1$ \\[25pt]


Sample Variance of Residual Faint Star Brightness & $\lambda I_{\lambda}^{\rm RS} \pm 0.6
                                 \, $\nw & $\pm 0.7$ \\[25pt]

DGL-$100 \, \mu$m Model Uncertainties & $\lambda I_{\lambda}^{\rm
                                        DGL}$$\{-8.7,+8.2\} \,$\nw & $\{-9.1,+8.6\}$ \\[10pt]

Galactic Extinction & $(1.05\pm0.01) \cdot \lambda I_{\lambda}^{\rm
  residual}$ & $-$ \\[5pt]

\hline
\multicolumn{3}{l}{$^{\ast}$ `$-$' denotes a negligible error.}
\end{tabularx}
\end{table}



\end{document}